\documentclass[prb,twocolumn,preprintnumbers,amsmath,amssymb]{revtex4}

\usepackage{graphicx}
\usepackage{dcolumn}
\usepackage{bm}
\usepackage{amsmath}
\usepackage{amssymb}
\usepackage{amsthm}
\usepackage{amsfonts}
\usepackage{enumerate}
\usepackage{latexsym}

\begin{document}

\title{Symmetry distinct spin liquid states and phase diagram of Kitaev-Hubbard model}

\author{Long Liang$^{1}$,  Ziqiang Wang$^{2}$, and Yue Yu$^{3,1}$}

\affiliation {${}^1$State Key Laboratory of Theoretical Physics, Institute of
Theoretical Physics, Chinese Academy of Sciences, P.O. Box 2735,
Beijing 100190, China\\
$^2$ Department of Physics, Boston College, Chestnut Hill,
  Massachusetts 02467, USA\\
${}^3$Department of Physics, Center for Field
Theory and Particle Physics, and State Key Laboratory of Surface Physics, Fudan University, Shanghai 200433,
China}

\date{\today}

\begin{abstract}
We report the finding of a series of symmetry distinct spin liquid (SL) states and a rich phase diagram in a half-filled honeycomb lattice Hubbard model with spin-dependent hopping amplitude $t'$.  We first study the magnetic instability of the system and find two antiferromagnetic (AF) orders beyond a  critical Hubbard $U$ which increases with the ratio $t'/t$ .   For $t'$ approaching to $t$, the semimetal (SM) transforms to a $U(1)$ SL and then to the Kitaev $Z_2$ SL as $U$ increases.  In a wide middle range of $t'/t$, the latter is replaced by a $U(1)$ SL to $SU(2)$ SL transition.   The physical properties of the stable SL phases are discussed.

\end{abstract}

\pacs{71.10.Fd, 71.27.+a, 71.30.+h}

\maketitle

\section{introduction}

Searching for the spin liquid (SL) states \cite{anderson,anderson1} in strongly correlated systems has been one of the most intriguing and important fields\cite{lee}.
Experimentally, possible SL states have been observed recently in quantum frustrated spin systems \cite{tria,hms,onh,iom,hhc,kago}. 
Theoretically, ample numerical evidence of SL ground state has been found in models of frustrated spin systems \cite{jws,yhw,jyb,dmu,jwb} and various exactly solvable models have been constructed to support the existence of the SL ground state \cite{FMS,kitaev2003,kitaev2006,lw2005,yk,yl}.

While most of the studies focused on quantum spin systems, possible SL states in the Hubbard model 
 near the metal-insulator transition have attracted great interests recently. Quantum Monte Carlo studies have produced controversial results \cite{mengsl,sorellansl,ah} amid an active debate over other
 investigations \cite{qc,qc1,qc2,qc3,qc4,ana1,ana2,sunkou}. Whether a SL ground state exists remains inconclusive due to the intricate interplay of charge and spin dynamics in the quantum critical region of the metal insulator transition and the incipient antiferromagnetic (AF) order.

We here study the ground state and the phase diagram of a generalized $t-t'-U$ Hubbard model where $t'$ describes nearest neighbor spin-dependent hopping 
on the half-filled honeycomb lattice. This model, which we call a Kitaev-Hubbard model, was introduced by Duan et al, and can be realized in cold atom systems \cite{duan,zhang,nil}, has been studied numerically at quarter filling\cite{hassan1} and half filling\cite{hassan2,hassan3}.
 Our motivation is that as a function of $t'/t$, this model interpolates between the usual Hubbard model at $t'/t=0$ and $1$ where the Kitaev $Z_2$ SL \cite{kitaev2006,zn} becomes the known ground state in the large on-site $U$ limit \cite{duan,JtJ,hassan2}. With the latter severing as a reference SL state, we obtain a rich phase diagram on the $U/t-t'/t$ plane that reveals the phase structure of several symmetry distinct SL states on the honeycomb lattice as well as their competition with several forms of AF order.

A SL is a Mott insulator without spin order that has exotic charge-neutral excitations called spinons which are coupled by emerging gauge fields\cite{wenbook}. Most commonly studied SLs are $U(1)$, $SU(2)$, and $Z_2$ SL - named after the symmetry of the gauge fields, respectively.  In this work, we show that these SL states emerge on the phase diagram of the Kitaev-Hubbard model at half filling and have yet to be explored by numerical studies \cite{hassan2}.
Specifically, we apply the $SU(2)$ slave rotor theory \cite{kim,hermele} to the Kitaev-Hubbard model, which recovers the semimetal (SM) phase with stable Dirac points at weak interactions, and search for all possible SL phases with gapless fermionic spinon excitations coupled to the corresponding gauge groups. We also calculate the spin susceptibility using the random phase approximation (RPA) and determine the stability of the SLs against magnetic ordering on the honeycomb lattice.
The physical characters of the different SLs that are experimentally distinguishable will be discussed.

The present paper is organized as follows. In section \ref{sec:model} we describe the model and its symmetries. We study the magnetic instabilities using RPA in section \ref{sec:RPA}. In section \ref{sec:sl}, we use $SU(2)$ slave rotor theory to study the SL phases. Then we describe how to combine the results of RPA and slave rotor to obtain the global  phase diagram in section \ref{sec:phase}. Finally, we discuss how to measure the different phases in cold atom experiments  in section \ref{sec:cd}. %a conclusion and discussion

\vspace{5pt}

\section{The model and its symmetries}\label{sec:model}

\begin{figure}
\begin{center}
\includegraphics[width=0.4\textwidth]{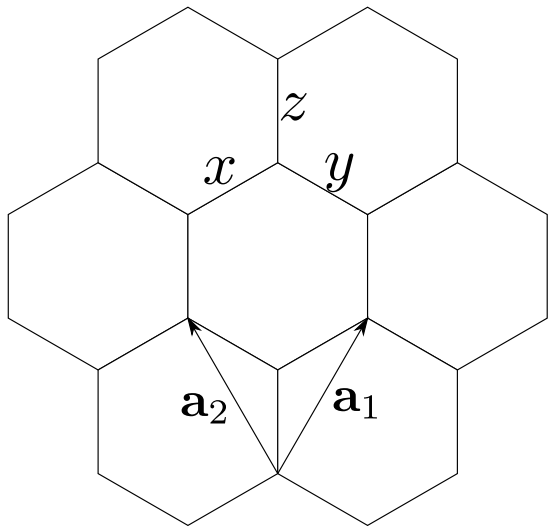}
\caption{Honeycomb lattice. $\mathbf{a}_1$ and $\mathbf{a}_{2}$ are primitive translation vectors. Nearest-neighbor links are divided into three types, called $x$-links, $y$-links and $z$-links.}
\label{fig:uc}
\end{center}
\end{figure}

%\noindent{\it The model and its symmetries. }
The Hamiltonian for the Kitaev-Hubbard model on a honeycomb lattice \cite{duan} is given by
\begin{eqnarray}
H=-t\sum_{\langle ij\rangle}c^{\dag}_{i}c_{j}-t'\sum_{\langle ij\rangle_{a}}c^{\dag}_{i}\sigma^{a}_{}c_{j}
+U\sum_{i}n_{i\uparrow}n_{i\downarrow},
\end{eqnarray}
where $c_{i}=[c_{i\uparrow},c_{i\downarrow}]^{T}$ and  $c_{i\sigma}$ ($c^{\dag}_{i\sigma}$) annihilates (creates) an electron with spin $\sigma=(\uparrow,\downarrow)$ on site $i$. The $t$-term is the conventional hopping integral, while the $t'$-term describes link-and-spin-dependent hopping, and $U$ is the on-site repulsion. $\sigma^a$ ($a=x,y,z$) are the Pauli matrices. $\langle ij\rangle_{a}$ denotes the nearest neighbor pairs in the $a$-link direction(see Fig.\ref{fig:uc}).

We group the electron operators into $2\times2$ matrices \cite{affleck1988}
\begin{eqnarray}
\Psi_{A}=\left[\begin{array}{cc}
c_{A\uparrow} & c^{\dag}_{A\downarrow} \\
c_{A\downarrow} & -c^{\dag}_{A\uparrow}
\end{array} \right],~~
 \Psi_{B}=\left[\begin{array}{cc}
c_{B\uparrow} & -c^{\dag}_{B\downarrow} \\
c_{B\downarrow} & c^{\dag}_{B\uparrow},
\end{array} \right]\label{rg}
\end{eqnarray}
where $A$ and $B$ label sites on the two sub-lattices, and define spin and pseudo-spin operators as  $$S^{a}=\frac{1}{4}\mathrm{Tr}{(\Psi^{\dag}\sigma^a\Psi)}, ~T^{a}=\frac{1}{4}\mathrm{Tr}{(\Psi\sigma^a\Psi^{\dag})}$$
$S^a$ acts on the subspace with odd electron number while $T^a$ acts on the subspace with even electron number so they are commutative with one  another.

To reveal the symmetry it's convenient to rewrite the Hamiltonian  in terms of $\Psi_A$ and $\Psi_B$:
\begin{eqnarray}\label{hamiltonian}
H&=&-t\sum_{\langle A,B\rangle}\mathrm{Tr}(\Psi^{\dag}_{A}\Psi_{B})-t'\sum_{\langle A,B\rangle_{a}}\mathrm{Tr}(\sigma^z\Psi^{\dag}_{A}\sigma^a\Psi_{B})\nonumber\\
&+&\frac{2U}{3}\sum_{i} \mathbf{T}_{i}^2.
\end{eqnarray}
The $t$-term clearly preserves spin and pseudo-spin rotational symmetries while the
$t'$-term also preserves spin and pseudo-spin rotational symmetries if we perform suitable local rotation of $\Psi$s. (See Appendix \ref{uc}.) However, when both $t$ and $t'$ are non-zero, both $SU(2)$ symmetries are broken since the symmetry operations of $t$-term and $t'$-term are not compatible with each other.
There is an important discrete chiral symmetry associated with the operator: $\mathcal{S}=\mathcal{P}\cdot\mathcal{T}$ where $\mathcal{P}=e^{i\pi T^y}$ is particle-hole and $\mathcal{T}=e^{i\pi S^y}K$ the time-reversal operation($K$ is complex conjugation). Because of this symmetry, TRS is enforced at half-filling in the large $U$ limit \cite{hassan2} as seen in the phase diagram in Fig.\ref{fig:PhD1}. In contrast to the usual lattice translation $T_{x}$ and $T_{y}$ and the inversion $\mathcal{R}$ symmetries that are preserved, the six-fold rotational symmetry is broken due to the $t'$-term. However, the system is invariant under combined $\pi/3$ lattice rotation and a spin rotation, 
$$\Psi_{A,B}\rightarrow e^{-i\frac{\pi}{4}\sigma^x}e^{-i\frac{\pi}{4}\sigma^{y}}\Psi_{A,B}$$ 

In the noninteracting limit ($U=0$) \cite{nil}, the electron dispersion $\varepsilon({\bf k})$ is given by
\begin{eqnarray}
 \varepsilon^{2}({\bf k})&=&t^2 |f|^2+3t'^{2}\pm[(t^2 |f|^2+3t'^{2})^2\nonumber\\
 &-&|t'^{2}(g_{+}g_{-}+1)-t^2f^2|^2]^{1/2},\label{fd}\end{eqnarray}
where $g_\pm=e^{ik_{1}}\pm ie^{ik_{2}}$ and $f({\bf k)}=1+e^{ik_{1}}+e^{ik_{2}}$.
$k_{1}=\mathbf{k}\cdot\mathbf{a}_1=k_{x}/2+\frac{\sqrt{3}k_{y}}{2}$, $k_{2}=\mathbf{k}\cdot\mathbf{a}_2=-k_{x}/2+\frac{\sqrt{3}k_{y}}{2}$, which maintains the original Dirac points on the honeycomb lattice at $(k_{x},k_{y})=(\pm 4\pi/3,0)$ where $g_{+}g_{-}=-1$ and $f=0$.
When $t'=0$, the dispersion near the Dirac point is $\varepsilon(k)=\pm \sqrt{3}t k/2$. When $t'\ne0$, the low energy physics is controlled by $t'$ and the dispersion near the Dirac points becomes $\varepsilon(k)=\pm t' k/2$.  Simple dimension counting shows that the Hubbard term is perturbatively irrelevant in the weak coupling limit.

\vspace{5pt}
\section{Magnetic instabilities and AF order}\label{sec:RPA}
%\noindent{\it Magnetic instabilities and AF order. }
 In this section we  determine the magnetic phase boundaries  by calculating the static spin susceptibility  $\chi(\mathbf{q})$ by RPA. This kind of calculation is standard\cite{nagaosabook} so we only present the main results here and the details can be found in Appendix \ref{rpa}.
 
\begin{figure}
\includegraphics[width=0.4\textwidth]{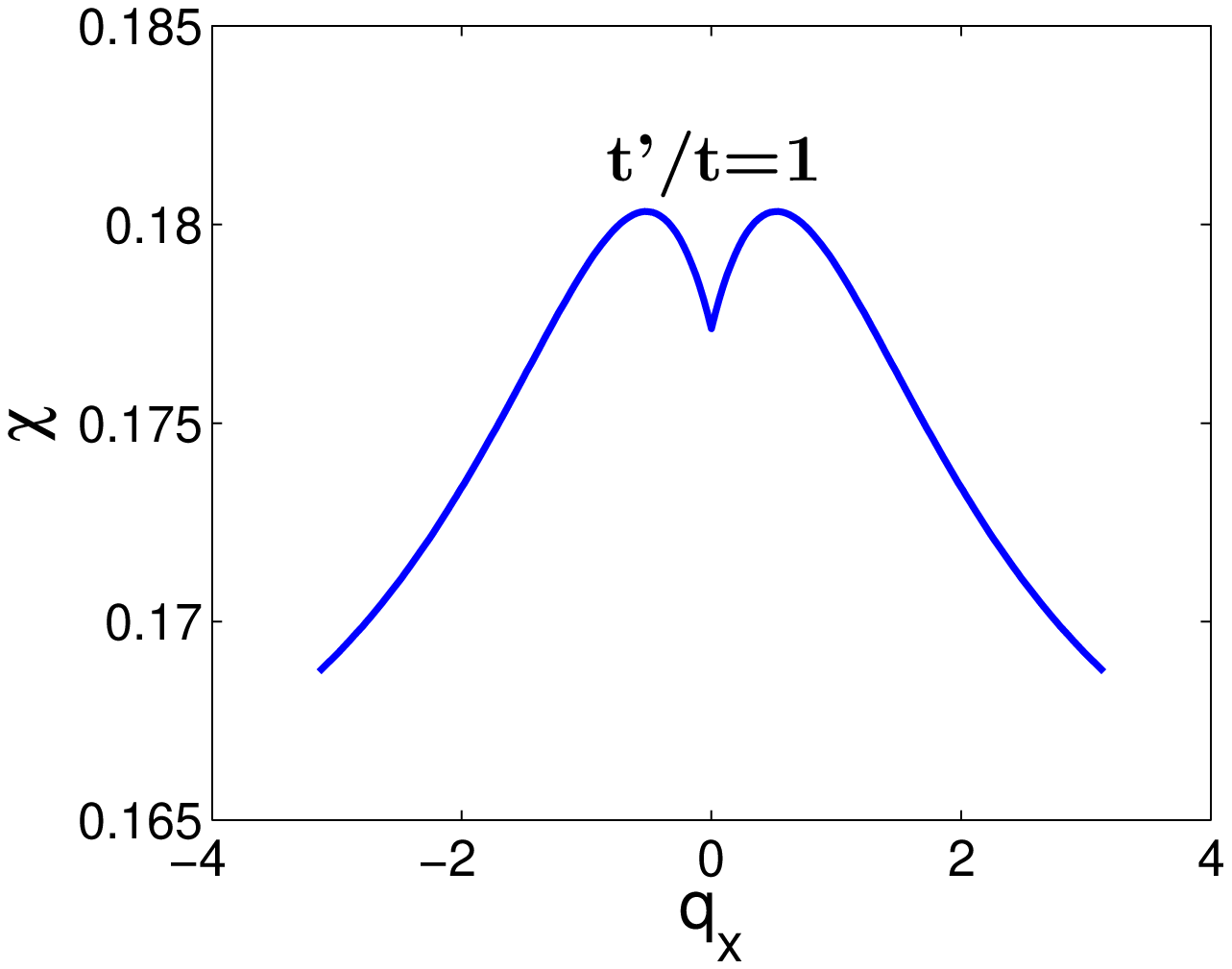}
\includegraphics[width=0.4\textwidth]{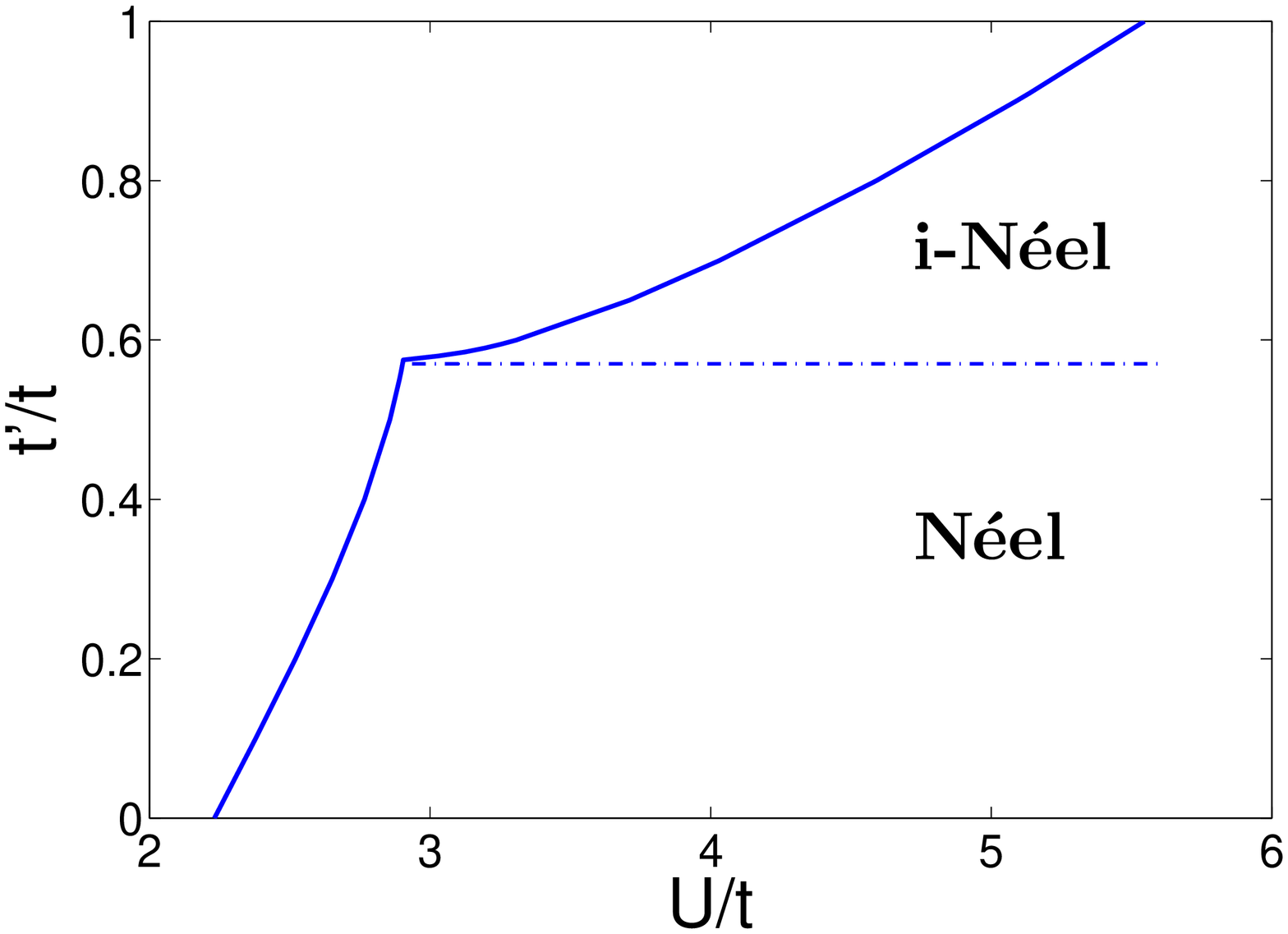}
%\centerline{(a)~~~~~~~~~~~~~~~~~~~~~~~~~~~~~~~~~~~~(b)}
\caption{(color online) Upper: An example of the double peaks of $\chi({\bf q})$ in i-N\'eel regime. Lower: i-N\'{e}el refers  to incommensurate AF order and there is no magnetic order in the blank area.}
\label{fig:afmrpa}
\end{figure}

We find that when $t'/t<0.57$, the peak in $\chi(\mathbf{q})$ is located at the $\Gamma$ point which means the magnetic phase is dominated by N\'eel order. For $t'/t>0.57$, the peak splits into two, indicative of the tendency toward incommensurate AF order. Fig.\ref{fig:afmrpa} (a) shows an example of the double-peak structure around the $\Gamma$ in the susceptibility in the incommensurate N\'eel (i-N\'eel) regime. The magnetic phase diagram is shown in Fig.\ref{fig:afmrpa} (b) where the SM and the magnetic phase boundary marks the onset of the divergence in the magnetic susceptibility $\chi(\mathbf{q})$ \cite{gf}. The appearance of the i-N\'eel AF regime has the same origin as that of the Kitaev SL phase due to the competition between the N\'eel order and the rotated N\'eel order \cite{JtJ}. Microscopically, this competition arises from the $t$- and $t'$-terms and pushes the magnetic phase boundary dramatically toward large $U$, enabling the possible stable SL phases in the enlarged nonmagnetic regime in Fig.\ref{fig:afmrpa} when $t'/t>0.57$. In the following, we will focus on the $t'/t>0.57$ regime and use the $SU(2)$ slave rotor theory to study the emergent SL phases.

\vspace{5pt}

\section{SU(2) slave rotor theory}\label{sec:sl}
%\noindent{\it SU(2) slave rotor theory.}

\subsection{A brief review of $SU(2)$ slave rotor theory}

Here we give a brief review of the $SU(2)$ slave rotor theory; more details can be found in\cite{kim,hermele}.
We start from Eq.\ref{hamiltonian} and decouple the Hubbard interaction by a Hubbard-Stratonovich (HS) transformation
$$\frac{i}{4} \sum_{i\in A,B}\mathrm{Tr}(\Psi_{i}\phi^{a}_{i}\sigma^a\Psi^{\dag}_{i})+\frac{3}{16U}\sum_{i\in A,B}\mathrm{Tr}(\phi^{a}_{i}\sigma^a)^2$$
 where  $\phi^{a}$ is a three-component HS field. To reveal the $SU(2)$ structure of the theory, we rotate the order parameters in the pseudo-spin space: $$\phi^{a}_{i}\sigma^a \rightarrow Z^{\dag}_{i}\phi^{a}_{i}\sigma^a Z_{i}$$
where $Z_{i}$ is a time-dependent $SU(2)$ matrix and can be parametrized as: $$Z_{i}=\left[\begin{array}{cc}
z_{i,1} & z_{i,2} \\
-z^{\ast}_{i,2} & z^{\ast}_{i,1}
\end{array}\right]$$ under the constraint $|z_{i,1}|^2+|z_{i,2}|^2=1$.
Such a rotation transforms the electron operator $\Psi_{i}$ to $F_{i}=\Psi_{i}Z^{\dag}_{i}$ but doesn't affect the spin operator ${\bf S}$. We thus call $F_{i}$ the spinon carrying the spin degrees of freedom and $Z_{i}$ the $SU(2)$ rotor tracking the pseudo-spin(charge) degrees of freedom. With a change of variable $$\phi^{a}_{i}\sigma^a\rightarrow\phi^{a}_{i}\sigma^a+2i Z_{i}\partial_{\tau}Z^{\dag}_{i}$$
the $SU(2)$ action reads
\begin{eqnarray}
S&=&S_t+\int^{\beta}_{0} \mathrm{d}\tau~\biggl\{\frac{1}{2}\sum_{i\in A,B}\mathrm{Tr}(F_{i}(\partial_{\tau}+\frac{i}{2}\phi^{a}_{i}\sigma^{a})F^{\dag}_{i})
\nonumber\\
&+&\frac{1}{2U'}\sum_{i}\mathrm{Tr}(\frac{1}{2}\phi^{a}_{i}\sigma^a+i Z_{i}\partial_{\tau}Z^{\dag}_{i})^2\label{rh}\\
&+&\sum_{i}i\lambda_{i}[\frac{1}{2}\mathrm{Tr}(Z^{\dag}_{i}Z_{i})-1]\biggr\},\nonumber
\end{eqnarray}
with
\begin{eqnarray}
S_t&=&\int_0^\beta \mathrm{d}\tau~\biggl\{-t\sum_{\langle A,B\rangle}\mathrm{Tr}(Z_{B}Z^{\dag}_{A}F^{\dag}_{A}F_{B})\nonumber\\
&-&t'\sum_{\langle A,B\rangle_{a}}\mathrm{Tr}(Z_{B}\sigma^z Z^{\dag}_{A}F^{\dag}_{A}\sigma^a F_{B})\biggr\}\label{t-action}
\end{eqnarray}
where $U'=2U/3$ and $\lambda_{i}$ is a local Lagrange multiplier imposing the $Z\in SU(2)$ constraint \cite{fandg}.

Eq.(\ref{rh}) describes a strongly coupled $SU(2)$ gauge theory, where $Z$ and $F$ are matter fields and $\phi$ are the temporal components of gauge fields. The gauge transformations are:
 $$F_{i}\rightarrow F_{i}W_{i},~Z_{i}\rightarrow W^{\dag}_{i}Z_{i},~ \phi^a_{i}\sigma^a\rightarrow\phi^a_{i}\sigma^a-2iW_{i}\partial_{\tau}W^{\dag}_{i}$$ 
The quartic matter fields in the hopping terms in Eq.(\ref{t-action}) can be further decoupled using the HS transformation, giving the spatial components of gauge fields.
Following Lee and Lee\cite{leelee} we decouple the hopping terms:

\begin{eqnarray}
S_{t}&=&\int^{\beta}_{0}\mathrm{d}\tau \biggl\{t\sum_{\langle A,B\rangle}\mathrm{Tr}(\eta_{AB}\eta^{\dag}_{AB})
-t\sum_{\langle A,B\rangle}\mathrm{Tr}(\eta_{AB}Z_{B}Z^{\dag}_{A})\nonumber\\
&-&t\sum_{\langle A,B\rangle}\mathrm{Tr}(\eta^{\dag}_{AB}F^{\dag}_{A}F_{B})+t'\sum_{\langle A,B\rangle_{a}}\mathrm{Tr}(\eta'_{AB}\eta'^{\dag}_{AB})\label{t-action2}\\
&-&t'\sum_{\langle A,B\rangle_{a}}[\mathrm{Tr}(\eta'_{AB}Z_{B}\sigma^z Z^{\dag}_{A})+\mathrm{Tr}(\eta'^{\dag}_{AB}F^{\dag}_{A}\sigma^a F_{B})]\biggr\}\nonumber
\end{eqnarray}
We can write\cite{hermele} $\eta_{AB}=|\eta|e^{i\theta_{AB}}e^{i(c^a_{AB}-id^{a}_{AB})\sigma^a}$ and $\eta'_{AB}=i|\eta'|e^{i\theta'_{AB}}e^{i(c^{'a}_{AB}-id^{'a}_{AB})\sigma^a}$. The action becomes complex for fluctuations of $\theta_{AB}$, $d_{AB}$, $\phi_i$ and $\lambda_{i}$. To get saddle point solutions with real free energy we perform analytic continuations, $\theta_{AB}\rightarrow i\tilde{\theta}_{AB}$, $d_{AB}\rightarrow i\tilde{d}_{AB}$, $i\phi_{i}\rightarrow\tilde{\phi}_{i}$ and $i\lambda_{i}\rightarrow\tilde{\lambda}_{i}$, where quantities with tildes are real. We can obtain 
$\eta_{AB}=|\eta_{Z}|e^{i(c^a_{AB}+\tilde{d}^{a}_{AB})\sigma^a}$, 
$\eta^{\dag}_{AB}=|\eta_{F}|e^{-i(c^a_{AB}-\tilde{d}^{a}_{AB})\sigma^a}$, 
$\eta'_{AB}=i|\eta'_{Z}|e^{i(c'^a_{AB}+\tilde{d}^{'a}_{AB})\sigma^a}$ and 
$\eta^{'\dag}_{AB}=-i|\eta'_{F}|e^{-i(c^{'a}_{AB}-\tilde{d}^{'a}_{AB})\sigma^a}$. $\eta/\eta'$ is a real/imaginary number times an
$SU(2)$ matrix.

Since we are interested in the half filling case, we will take $\tilde{\phi}^{a}_{i}=0$ ans\"atz for the temporal components of the gauge fields and relax the constraint by setting $\lambda_{i}=\lambda$, as is always done in slave particle theories. 
Despite the large gauge fluctuations, it is possible to obtain deconfined phases by different mean field ans\"atz \cite{lnw}.

\vspace{5pt}

\subsection{$Z_2$ and $SU(2)$ SL phases}
%\noindent{\it $Z_2$ and $SU(2)$ SL phases. }
The generic ans\"atz,
\begin{eqnarray}
\small \eta^{\dag}_{AB}=a_{F}\sigma^{0}, \eta'^{\dag}_{AB,a}=a'_{F}\sigma^{a}, \eta_{AB}=a_{Z}\sigma^{0}, \eta'_{AB,a}=a'_{Z}\sigma^{a}, \nonumber
\end{eqnarray}
breaks the $SU(2)$ gauge symmetry to $Z_{2}$, where the subscript $a=x, y, z$ denotes the bond type.
It preserves TRS and is not thus valid in the  weak coupling.
Denoting $t_{\uparrow}=-ta_{F}f-t'a'_{F}$, $t_{\downarrow}=-ta_{F}f+t'a'_{F}$, $\Delta_{\uparrow}=-t' a'_{F}(e^{i k_1}+e^{i k_2})$, and $\Delta_{\downarrow}=-t' a'_{F}(e^{i k_1}-e^{i k_2})$, the dispersion of the spinon is given by
\begin{eqnarray}
\varepsilon_{F}=\pm\frac{1}{2}|t_{\sigma}\pm\Delta_{\sigma}|,
\end{eqnarray}
which is the same as that obtained in \cite{bn, ybk}. The spinon band structure in the $Z_2$ phase is depicted in Fig. \ref{fig:spinon} (Upper panel). 
 mean field equations are derived in Appendix \ref{rotor}.
There are two critical lines in the mean field solutions.  For $U<U_{c1}$, the solution gives an improper $p$-wave superfluid phase that does not recover the SM when $U\to 0$. When $U>U_{c1}$, rotors are gapped and the $Z_2$ SL arises. Remarkably, a second critical line $ U_{c2}$ exists for $t'/t\lesssim 0.91$, such that when $U> U_{c2}$, $a'_{F}$ and $a'_{Z}$ vanish, i.e., the spin dependent hopping renormalizes to zero. The $SU(2)$ gauge symmetry is thus restored and the system enters an $SU(2)$ SL phase.

\begin{figure}[htb]
\centering
\includegraphics[width=6.5cm]{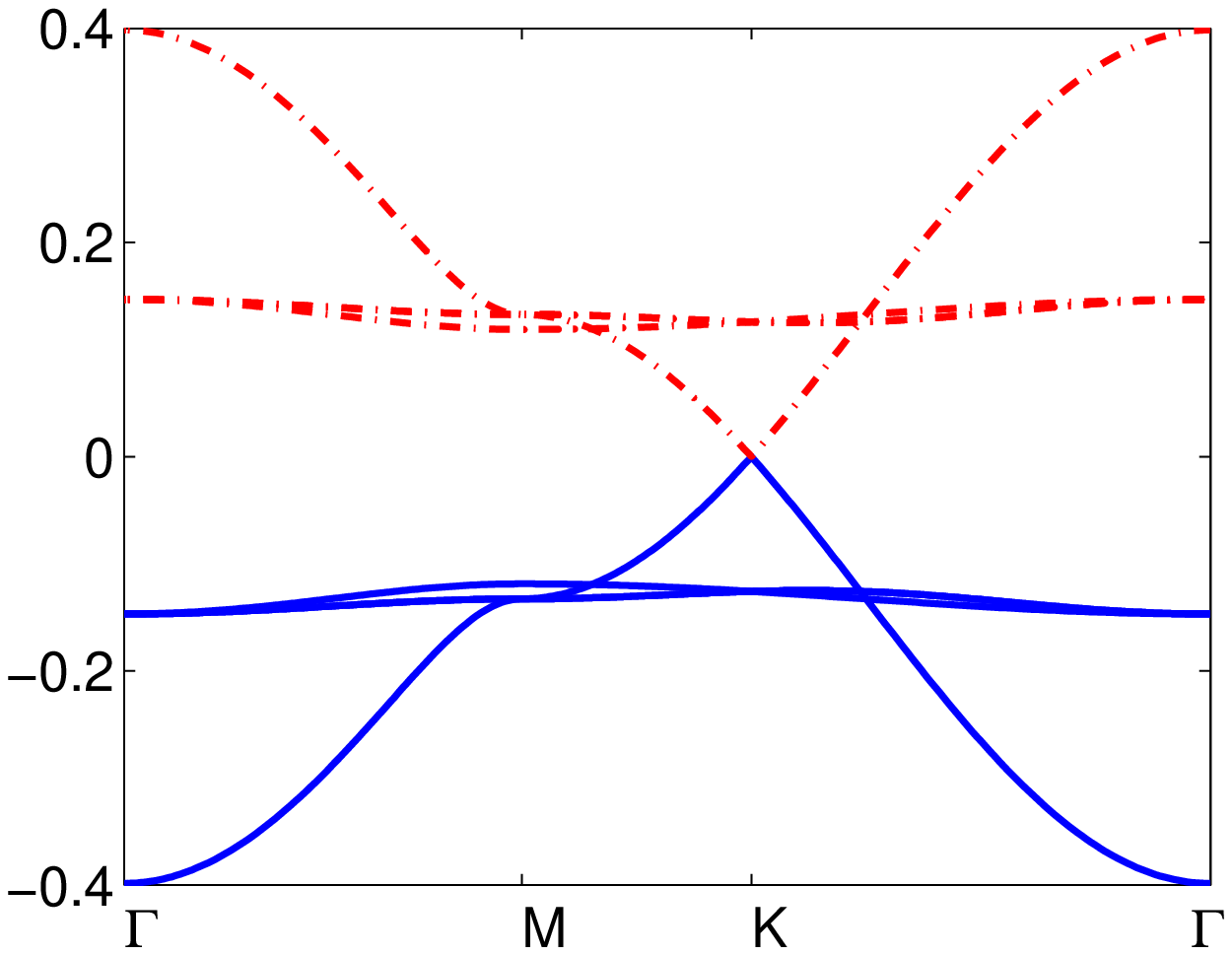}
\includegraphics[width=6.5cm]{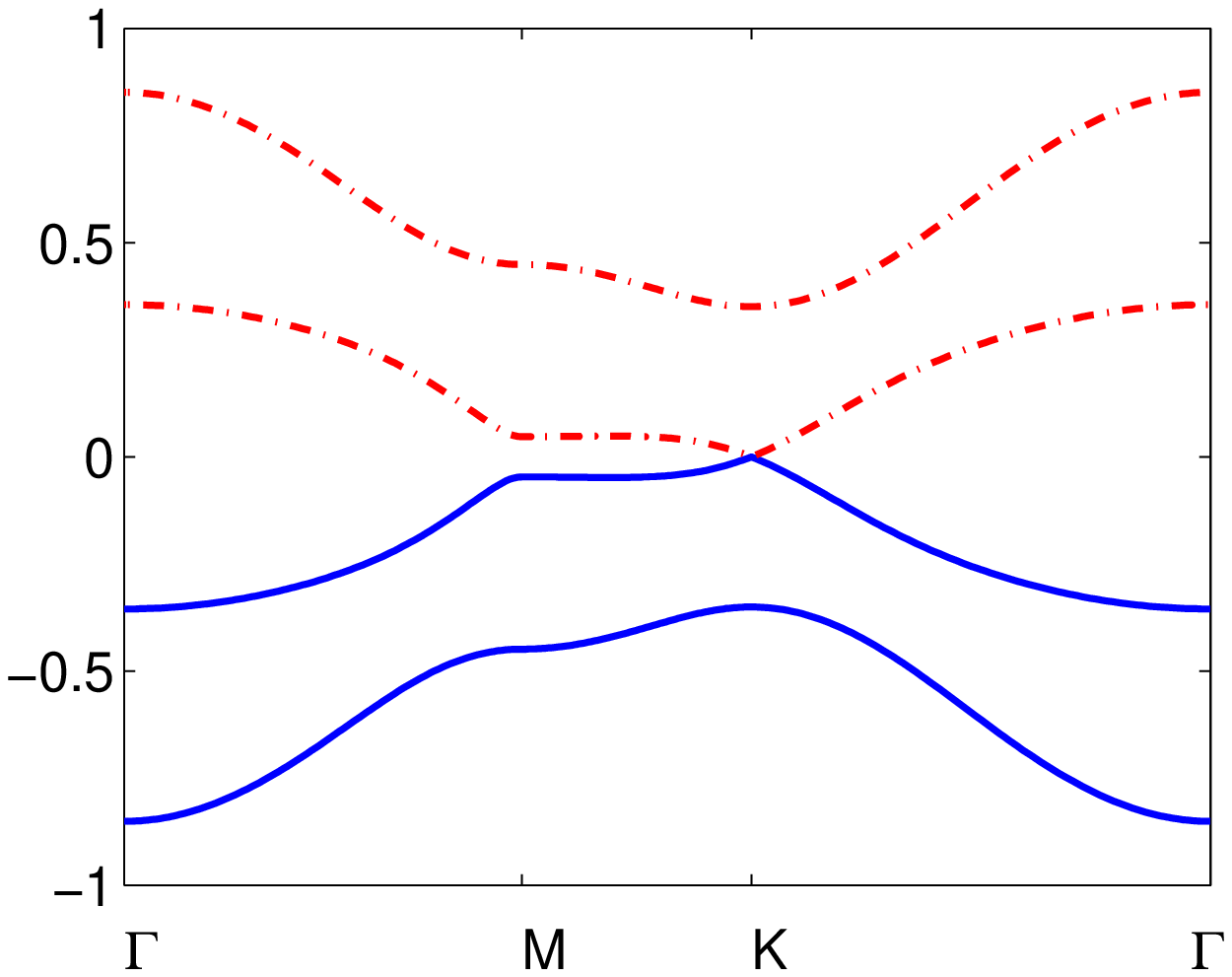}
\caption{(color online) Upper: Spinon band structure in $Z_2$ SL phase. $U=4t$ and $t'=t$. 
Lower: Spinon band structure in $U(1)$ SL phase. $U=3.67t$ and $t'=0.9t$.
Bands shown in blue solid line are occupied, in red dash line are unoccupied.
Spinon dispersion in $SU(2)$ phase looks like the one in  $Z_2$ phase with no (nearly) flat band.}
\label{fig:spinon}
\end{figure}
\vspace{5pt}

It is instructive to study the $Z_{2}$ ans\"atz for large-$U$ more carefully and compare to exact results at $t'/t=1$. The effective chemical potential of rotors is now $\lambda=U'=2U/3$ and the
$Z_2$ mean field equations are given by
\begin{eqnarray}\label{deta}
&&a_{Z}=\frac{1}{6tN}\sum_{\mathbf{k}}
\frac{\partial \varepsilon_{F,\sigma\pm}}{\partial a_{F}},~~ a'_{Z}=\frac{1}{6t'N}\sum_{\mathbf{k}}
\frac{\partial \varepsilon_{F,\sigma\pm}}{\partial a'_{F}} \\
&&
a_{F}=\frac{1}{24U'tN}\sum_{\mathbf{k}}
\frac{\partial \varepsilon^2_{Z,\pm}}{\partial a_{Z}},~~
a'_{F}=\frac{1}{24U't'N}\sum_{\mathbf{k}}
\frac{\partial \varepsilon^2_{Z,\pm}}{\partial a'_{Z}}\nonumber,
\end{eqnarray}
where $\varepsilon_{F\sigma\pm}$ and $\varepsilon_{Z,\pm}$ are the spinon and the rotor dispersions respectively. $\varepsilon_{F\sigma\pm}$ gives Majorana fermion excitations at the Dirac point and six gapped flat bands.
Note that $\varepsilon_{Z,\pm}\sim t$ and then $a_{F},  a'_{F}\sim t/U$, as expected for the charge and spin excitations in this limit.   When $t=t'$, the velocity of the  linear spinon dispersion in the Kitaev model, i.e., in the large $U$ limit, is determined by $J=\frac{8t^2}{U}$. In order to be consistent with this velocity, the parameter $a_{F}$ in the mean field spinon dispersion needs to be rescaled to $a_F=\frac{J}{16t}$, which amounts to rescale the Hubbard $U$ by a factor $\alpha\approx0.572$ at $t'/t=1$ (see Appendix \ref{rotor}). Note that it is well-known that the Hubbard $U$ needs to be rescaled in the mean field approximation of the slave rotor theory \cite{fandg}.  We would like to point out that the rescale of $U$ only affects the results quantitatively. We will demonstrate below that $\alpha$ is essentially independent of $t'/t$ in the regime where the slave rotor theory can be considered reliable. We solve the mean field equations self-consistently with the rescaled $U$. 

\vspace{5pt}

\subsection{$U(1)$ SL phases}
%\noindent{\it $U(1)$ SL phases. }
 To recover the semi-metal phase in small $U$ regime, we consider the following ans\"atz,
 \begin{eqnarray}
 \eta^{\dag}_{AB}=a^0_F\sigma^0, \eta'^{\dag}_{AB,a}=a^z_F\sigma^z,\eta_{AB}=a^0_Z\sigma^0, \eta'_{AB,a}=a^z_Z\sigma^z,\nonumber
 \end{eqnarray}
which breaks the $SU(2)$ gauge symmetry to $U(1)$ in general. The dispersion of the spinon is thus given by
\begin{eqnarray}
\varepsilon^{2}_{F}&=&t^2a^{02}_{F} |f|^2+3t^{'2}a^{z2}_{F}\pm\bigl[(t^2a^{02}_{F} |f|^2+3t'^{2}a^{z2}_{F})^2
\nonumber\\&-&|t'^{2}a^{z2}_{F}(g_{+}g_{-}+1)-t^2a^{02}_{F}f^2|^2\bigr]^{1/2}, \label{disp}
\end{eqnarray}
which has the same form as Eq.~(\ref{fd}) with renormalized hoppings $t\to ta^0_F$ and $t'\to t'a^z_F$.
The spinon band structure in the $U(1)$ phase is depicted in Fig. \ref{fig:spinon} (Lower panel). 
Hence Hence, we expect it to be favored near the weak coupling where the TRS is broken.
At the Dirac points, the linear dispersion $\varepsilon_F=\pm t'|a^z_F|k/2$. We again obtain two critical lines $\tilde U_{c1}$ and $\tilde U_{c2}>\tilde U_{c1}$. When $U<\tilde U_{c1}$, the rotors condense and the system is in the weak coupling SM phase. For $U>\tilde U_{c1}$, the rotors are gapped and the $U(1)$ SL phase emerges. For $U>\tilde U_{c2}$, both $a^z_{F}$ and $a^z_{Z}$ vanish and the system enters the $SU(2)$ SL phase.

\vspace{5pt}

\begin{figure}[htb]
\centering
\includegraphics[width=6.5cm]{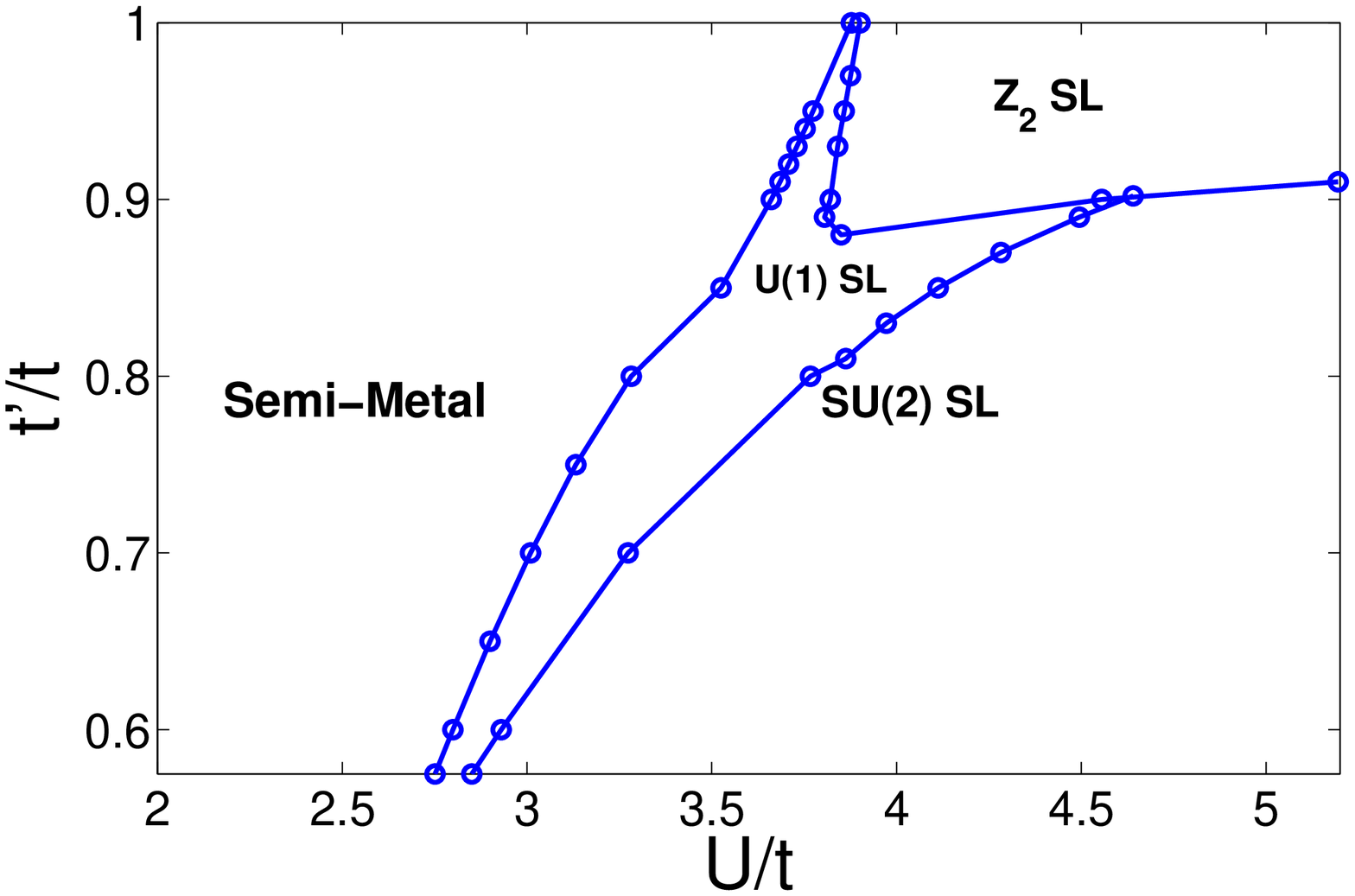}
\caption{(color online) Phase diagram of the SL states in the absence of magnetic order. Here we take $\alpha=0.572$.}
\label{fig:PhDMF}
\end{figure}
\vspace{5pt}

At a given point in the $U/t$ - $t'/t$  plane, which SL state is favored can be determined by comparing the mean field ground state energies among the $U(1)$, $Z_{2}$ and $SU(2)$ ans\"atz.  The obtained slave-rotor phase diagram is shown in Fig.\ref{fig:PhDMF}. 

%\subsection{Stability of SLs }
%\noindent {\it Determination of the global phase diagram. } 

Before ending this section, we discuss the stability of the SL states. To determine whether the SLs are stable one have to go beyond the mean field theory and consider the gauge fluctuations. We briefly discuss this issue here. The $Z_{2}$ spin liquid is stable because the gauge fluctuations are gapped. 
The $SU(2)$ spin liquid is the one study by Hermele\cite{hermele}, in this phase the low energy effective theory is gapless Dirac fermion coupled to compact $SU(2)$ gauge fields. Large $N$ expansion shows that when the number of fermion flavors is large enough, this spin liquid phase is stable\cite{hermele}. The effective theory of $U(1)$ spin liquid is gapless Dirac fermions coupled to compact $U(1)$ gauge field, this phase may also be stable against instanton effect\cite{hermele,hermele2004,sslee2008}.

\vspace{5pt}

\section{Determination of the global phase diagram}\label{sec:phase}
\begin{figure}[htb]
\begin{center}
\includegraphics[width=7cm]{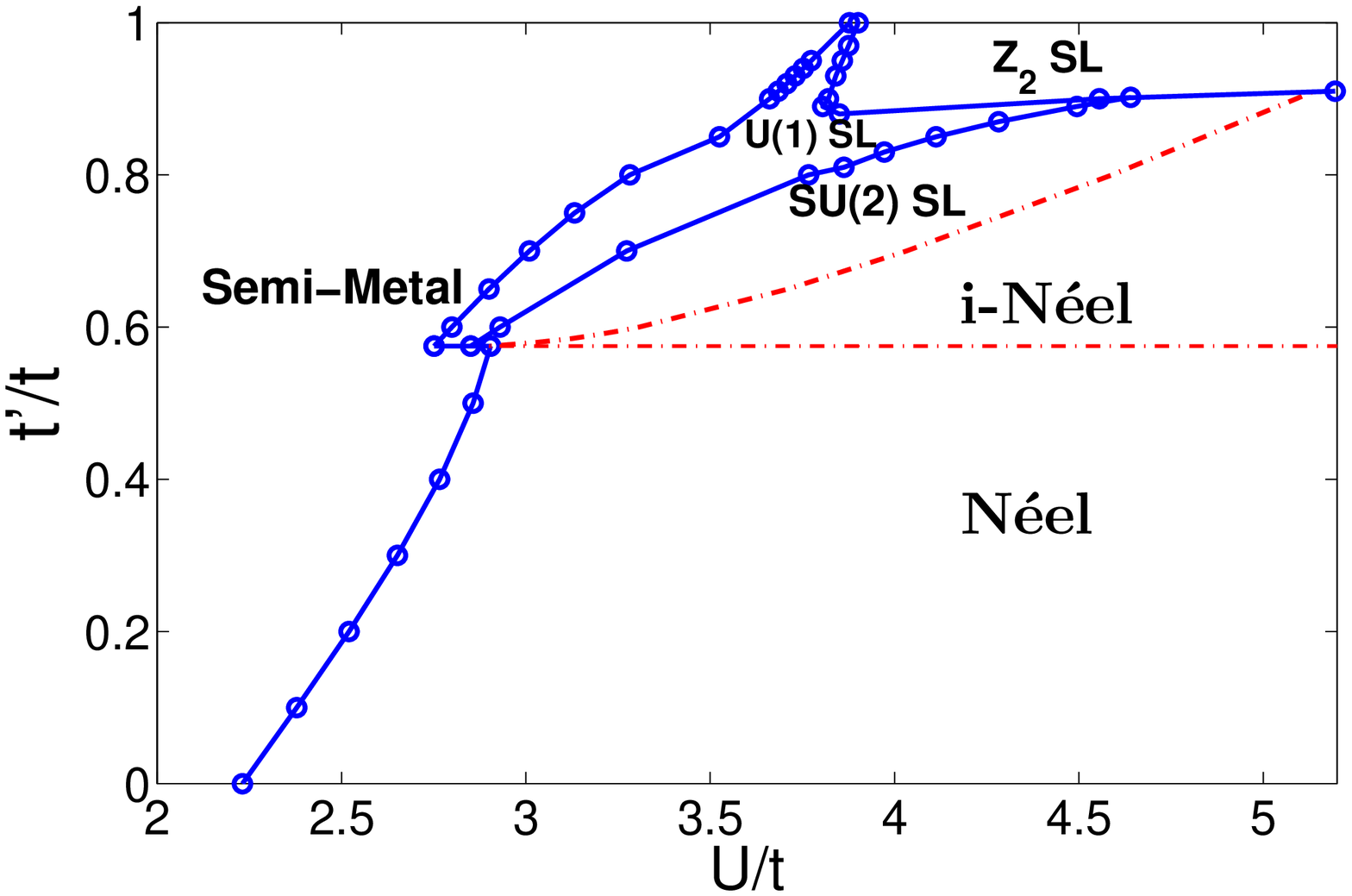}
\end{center}
\caption{\label{fig:PhD1} (color online) Phase diagram. The magnetism part comes from Fig. \ref{fig:afmrpa}(b) and the SL part from Fig. \ref{fig:PhDMF}. We use the red dash-dot lines to separate the SL and AF  phases as well as i-Neel and Neel because the RPA calculation in the strong coupling  is not as good as that in the weak coupling. }
\end{figure}

%\noindent {\it Determination of the global phase diagram. } 
 %The stability of SL states has been discussed in the literature \cite{hermele,hermele2004,sslee2008,hassan2}. 
For the Kitaev-Hubbard model, in particular, a numerical calculation based on a variational cluster approximation and cluster perturbation theory showed that the SL phase is unstable against the SM and AF states when $t'/t$ is smaller than a certain value\cite{hassan2}.  We now explain how to combine the SL phase diagram in Fig.\ref{fig:PhDMF} and the magnetic phase diagram in Fig. \ref{fig:afmrpa} to arrive at the global phase diagram shown in Fig.\ref{fig:PhD1}.
First, it is known when $t'/t\sim 1$ the slave rotor theory is reliable, while the RPA theory
provides the leading magnetic 
instability for all  $t'/t$.
The numerical work in Ref.\cite{hassan2} showed that there is a tricritical point for the SL, SM and AF phases. This tricritical point in our result corresponds to the discontinuity point in the slope of the phase boundary in Fig. \ref{fig:afmrpa}(b).
Remarkably, we find that the lower bound of the SL phase touches the singular point of the magnetic phase boundary, forming the tricritical point observed by numerical simulations \cite{hassan2}. We stress that the tricritical point emerges in our theory without the need to change the rescaling factor $\alpha$ determined by the exact solution of the Kitaev model at $t^\prime/t=1$ and lends further support for an essentially $t^\prime/t$-independent $\alpha$ in the SL regime. We emphasize that the topology of the SL phase diagram is not affected by varying $\alpha$ (see Appendix \ref{asl}). However, the tricritical point exists only if  $\alpha\approx 0.572$.
With increasing $U$, the $SU(2)$ SL
becomes unstable to the AF i-N\'eel phase. However, the AF phase terminates when it meets the $Z_2$ SL because the Kitaev SL has lower energy. 
 Finally, we would like to remark that although the general phase structure of our theory captures that of the numerical results with unprecedented symmetry distinct SL phases, the phase boundaries between the SM, AF and SL phases as well as the tricritical point only qualitatively agree with the numerical results in Refs. \cite{mengsl, sorellansl, hassan2}. The exact determination of the phase boundaries is beyond the scope of the current work.

%\noindent{\it Phase diagram. } 
It's time to describe our main results shown in the phase diagram shown in Fig.\ref{fig:PhD1}. Generally speaking, we found three types of phases: The SM phase for weak coupling, the AF phase for strong coupling $t'/t< 0.91$, and several SL states in-between. There are crucial differences between these new findings and the previous analytical and numerical results \cite{duan,JtJ,hassan2}.
(i) The presence of three types of symmetry distinct gapless $U(1)$, $SU(2)$, and $Z_2$ SL phases that are experimentally distinguishable. While the $Z_2$ SL encloses the exact solvable Kitaev spin model \cite{kitaev2006} at $t'/t=1$ and $t/U\ll 1$ and preserves the time-reversal symmetry (TRS), we discovered a TRS-breaking $U(1)$ SL that separates the $Z_2$ SL from the SM for $t'/t>0.91$.
The spinon dispersion in the $U(1)$ SL has the same form as that of the quasiparticle in the SM phase but with renormalized hoppings.
For $t'/t<0.91$, the $U(1)$ SL transforms with increasing $U$ into the $SU(2)$ SL
 where a free Dirac fermion spinon dispersion arises with $t'$ renormalized to zero and the TRS restored.
(ii) Our RPA results captures qualitatively several different AF ordered phases in different parameter regions. The presence of the link-spin dependent hopping $t^\prime$ introduces a competition between conventional N\'eel order and a new type of AF order accompanied by a local spin rotation \cite{JtJ}. The latter pushes  the magnetic phase boundary toward larger-U dramatically when $t'/t>0.57$, realizing the various symmetry distinct SLs as stable phases of the electronic matter.

\section{Experimental Implications}\label{sec:cd}
%\noindent{\it Conclusions and discussions.} 

To sum up, we studied a Kitaev-Hubbard model using RPA and slave rotor theory. We obtained a fruitful phase diagram, including semi-metal phase, commensurate and incommensurate AFM ordered phases and three symmetry distinct SL states.
We now discuss how to measure these phases in cold atom experiments.

%The anti-ferromagnet order can be measured via optical Bragg scattering\cite{braggaf}

If the SLs proposed are stable, they may be recognized in cold atom experiments. For example, 
Bragg spectroscopy can be used to measure the full band structure (see Fig.\ref{fig:spinon}) in cold atoms system\cite{braggs}.
There are some 
qualitative difference of the spinon dispersions in the $Z_2$, $U(1)$ and $SU(2)$ SL phases. 
The $Z_2$ SL differs apparently from the other two because there are Majorana fermion excitations and non-abelian anyonic Majorana bound states in an external magnetic field \cite{kitaev2006}.
The $U(1)$ spinon is of a linear dispersion proportional to $t'|a_F^z|$ at the Dirac points and does not have a conserved $S^z$, the dispersion of the $SU(2)$ spinon is the same as that of the free Dirac fermion with a conserved $S^z$ and a renormalized hopping $ta_F^0$.  

\begin{figure}[htb]
\centering
\includegraphics[width=6.5cm]{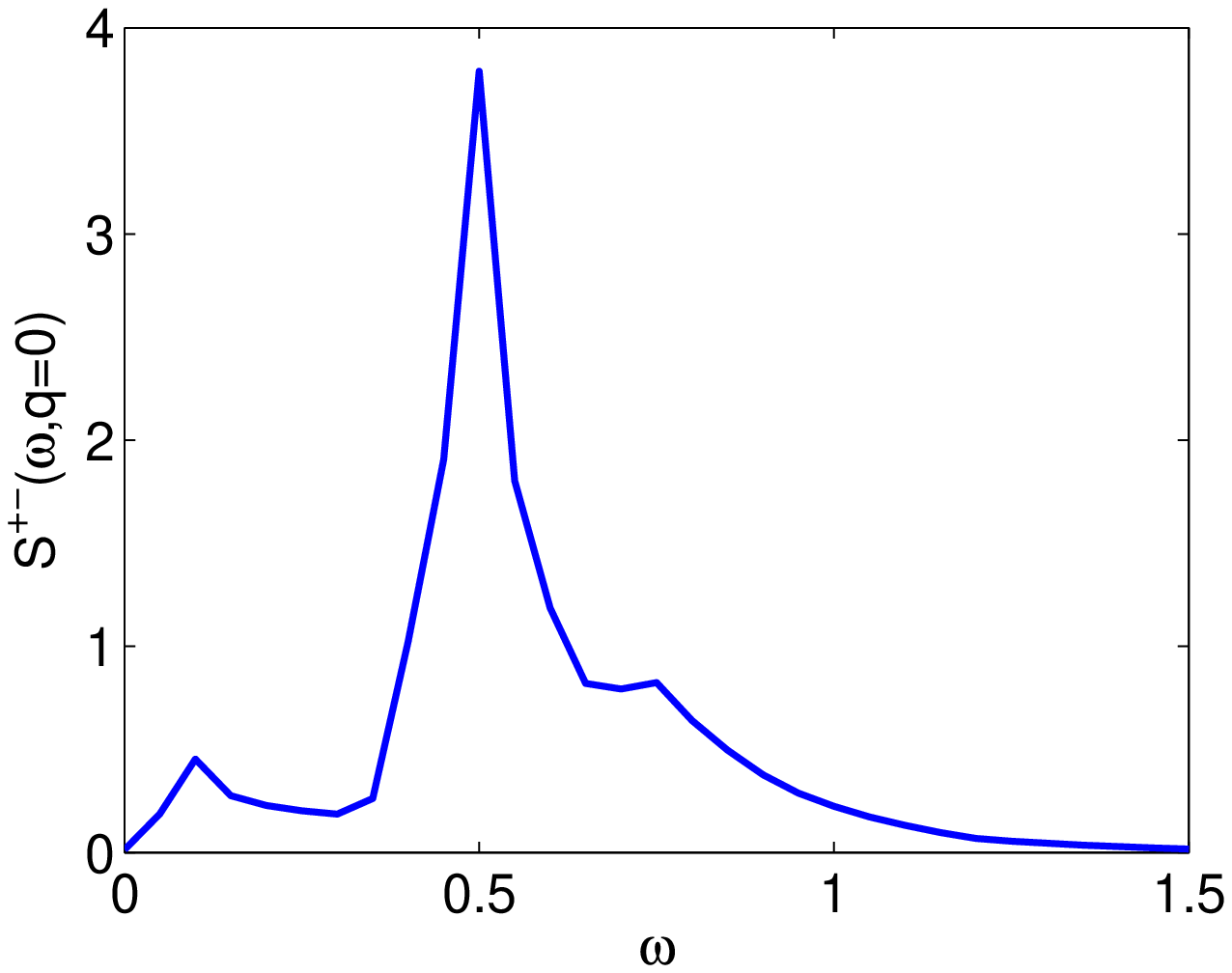}
\includegraphics[width=6.5cm]{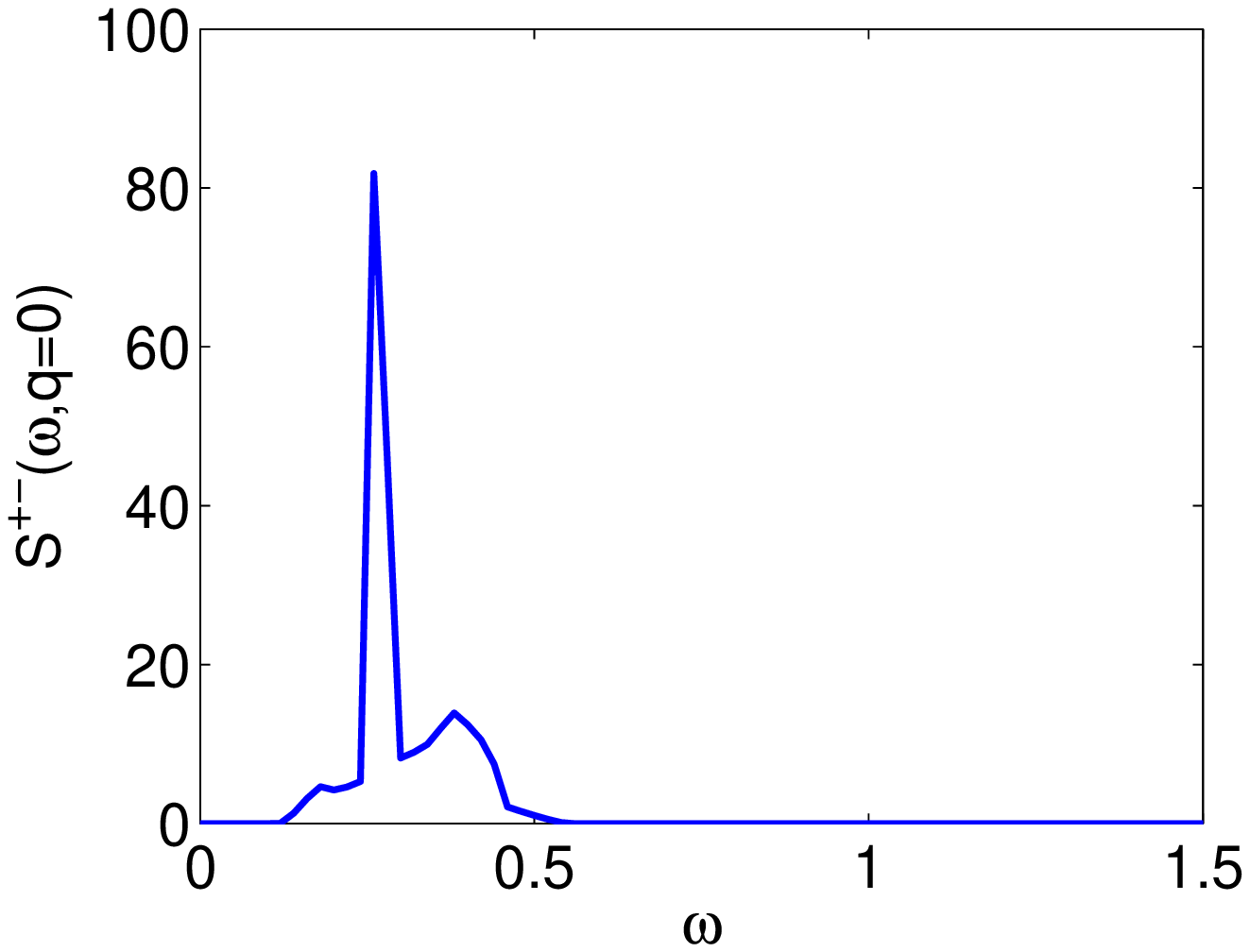}
\caption{(color online) Upper: Dynamical spin structure factor in $U(1)$ spin liquid phase. $U=3.67t$ and $t'=0.9t$. Lower: Dynamical spin structure factor in $Z_2$ spin liquid phase. $U=4t$ and $t'=t$}
\label{fig:dssf}
\end{figure}

Bragg spectroscopy can also be used to determine the dynamical spin structure $S^{+-}(\omega,\mathbf{q})$, which is the Fourier transformation of spin-spin correlation $\langle S^+(\mathbf{r},t)S^-(\mathbf{r}',t') \rangle$ and proportional to the cross section of Bragg scattering\cite{braggc} . In the $U(1)$ SL phase, because of the spin flip terms in the effective spinon Hamiltonian, $S^{+-}(\omega,\mathbf{q}=0)\ne 0$. In the small $\omega$ limit the cross section is proportional to the density of states near the Fermi surface (Dirac points), so $S^{+-}(\omega,\mathbf{q}=0)\propto\omega$ for small $\omega$ (see Fig.\ref{fig:dssf} Upper panel). In the $Z_2$ spin liquid phase, $S^{+-}(\omega,\mathbf{q}=0)=0$ if $\omega$ is smaller than the gap of the (nearly) flat band and a sharp peak appears when the energy transfer is twice the gap (see Fig.\ref{fig:dssf} Lower panel).  
In contrast, $S^{+-}(\omega,\mathbf{q}=0)=0$ in $SU(2)$ SL phase. These properties can be used to distinguish the SLs. 

The anti-ferromagnet order can also be measured via Bragg scattering\cite{braggaf}.

\vspace{5pt}

\centerline{Acknowledgement}

The authors thank Sen Zhou for useful discussions. This work is supported by the 973 program of MOST of China (2012CB821402), NNSF of China (11174298, 11121403), DOE grant DE-FG02-99ER45747 and NSF DMR-0704545. ZW thanks Aspen Center for Physics for hospitality.

\appendix

\section{Symmetries of $t'$-term}\label{uc}

If $\Psi_A$ is redefined as 
$$
\Psi_{A}=\left[\begin{array}{cc}
c_{A\uparrow} & -c^{\dag}_{A\downarrow} \\
c_{A\downarrow} & c^{\dag}_{A\uparrow}
\end{array} \right]$$
then the $t'$-term can be written as $$-t'\sum_{\langle A,B\rangle_{a}}\mathrm{Tr}(\Psi^{\dag}_{A}\sigma^a\Psi_{B})$$
 which is  pseudo-spin rotational invariant.

\begin{figure}[htb]
\begin{center}
\includegraphics[width=.4\textwidth]{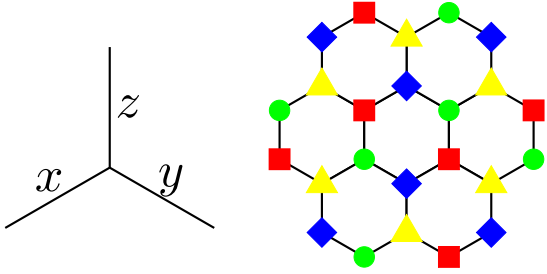}
\caption{(color online) $t'$-term can be written in a spin rotational invariant manner if the unit cell is enlarged.}
\label{fig:euc}
\end{center}
\end{figure}

To reveal the spin-rotational symmetry of 
$t'$-term,  one can enlarge the unit cell and perform local spin rotations of $\Psi$s (see Fig.\ref{fig:euc}): for circle , $\Psi\rightarrow\Psi$, for square, $\Psi\rightarrow\sigma^z\Psi$, for diamond, $\Psi\rightarrow\sigma^y\Psi$, for triangle, $\Psi\rightarrow\sigma^x\Psi$. After this rotation, $t'$-term can can be written in spin-rotational invariant way. However, even after this rotation, $t'$-term can't be written as the same form as $t$-term. Because for $t'$-term, electron acquire $\pi$ phase when hopping around a hexagon and this phase can not be removed by spin rotations.

Note that the symmetry operations of $t$-term and $t'$-term are not compatible, that's to say, $t$-term is invariant under some operations while $t'$-term is invariant under others, so both symmetries are broken when $t$ and $t'$ are nonzero.

\section{Calculation  of spin susceptibility}\label{rpa}

The partition function is 
$Z=\int Df^{\dag}D f e^{-\int^{\beta}_{0} L d\tau}$, 
where
$$L=\sum f^{\dag}_{i\sigma}(\partial_{\tau}\delta_{ij}\delta_{\sigma\sigma'}-t^{\sigma\sigma'}_{ij})f_{j\sigma'}+U\sum n_{i\uparrow}n_{i\downarrow}$$

After performing a Hubbard-Stratonovich transformation in spin channel, we get:
$Z=\int D f^{\dag} Df D\phi e^{-S}$
and
\begin{widetext}
$$S=\int^{\beta}_{0} d\tau \sum_{i}f^{\dag}_{i}\partial_{\tau}f_{i}
-t\sum_{\langle ij\rangle_{a}}f^{\dag}_{i}(I+\sigma^{a})f_{j}+\frac{U}{4}\sum_{i}\phi^{2}_{i}
+\frac{U}{2}\sum_{i}\phi_{i}f^{\dag}_{i}\sigma^zf_{i}$$
\end{widetext}
Integrating out fermions we get the effective action 
%\begin{widetext}
\begin{eqnarray}
S_{\mathrm{eff}}=\frac{U}{4}\int^{\beta}_{0}d\tau \sum\phi_{i}(\tau)^2
-\mathrm{Tr}\mathrm{ln}[\partial_{\tau}-t^{\sigma\sigma'}_{i,j}+\frac{U}{2}\sigma\phi_{i}
]\nonumber
\end{eqnarray}
%\end{widetext}
Setting $\phi_{A}=-\phi_{B}=\phi$, then up to second order and in static limit:
$S_{\mathrm{eff}}=\sum_{\mathbf{q}}\frac{U}{4}[1-U\chi(\mathbf{q})]\phi(\mathbf{q})
\phi(-\mathbf{q})$
where 
$\chi(\mathbf{q})=-\frac{1}{2\beta N}\sum_{\mathbf{k},\omega_{n}}\mathrm{Tr}G_{0}(i\omega_{n},\mathbf{k})\sigma^z\otimes\tau^z
G_{0}(i\omega_{n},\mathbf{k}+\mathbf{q})\sigma^z\otimes\tau^z$ and $G_{0}(i\omega_{n},\mathbf{k})=\frac{1}{i\omega_{n}-\mathcal{H}(\mathbf{k})}$ is bare Green's function.
If $\mathcal{H}$ is diagonalized by a matrix $V$, i.e., $V^{\dag}(\mathbf{k})\mathcal{H}(\mathbf{k})V(\mathbf{k})=diag(E_{h \mathbf{k}},-E_{h \mathbf{k}},E_{l \mathbf{k}},-E_{l \mathbf{k}})$, then

\begin{widetext}
\begin{eqnarray}\label{chi}
\chi(\mathbf{q})&=&\frac{1}{2N}\sum_{\mathbf{k}}
\frac{W_{12\mathbf{k,q}}W_{21\mathbf{k,-q}}}{E_{h\mathbf{k-q/2}}+E_{h\mathbf{k+q/2}}}+
\frac{W_{14\mathbf{k,q}}W_{41\mathbf{k,-q}}}{E_{h\mathbf{k-q/2}}+E_{l\mathbf{k+q/2}}}
+\frac{W_{21\mathbf{k,q}}W_{12\mathbf{k,-q}}}{E_{h\mathbf{k+q/2}}+E_{h\mathbf{k-q/2}}}+
\frac{W_{23\mathbf{k,q}}W_{32\mathbf{k,-q}}
}{E_{l\mathbf{k+q/q}}+E_{h\mathbf{k-q/2}}}\nonumber\\
&+&\frac{1}{2N}\sum_{\mathbf{k}}
\frac{W_{32\mathbf{k,q}}W_{23\mathbf{k,-q}}}
{E_{l\mathbf{k-q/2}}+E_{h\mathbf{k+q/2}}}+
\frac{W_{34\mathbf{k,q}}W_{43\mathbf{k,-q}}}{E_{l\mathbf{k-q/2}}+E_{l\mathbf{k+q/2}}}
+\frac{W_{41\mathbf{k,q}}W_{14\mathbf{k,-q}}}{E_{l\mathbf{k-q/2}}+E_{h\mathbf{k+q/2}}}+
\frac{W_{43\mathbf{k,q}}W_{34\mathbf{k,-q}}}{E_{l\mathbf{k+q/2}}+E_{l\mathbf{k-q/2}}}
\end{eqnarray}
\end{widetext}
where $W(\mathbf{k,q})=V^{\dag}(\mathbf{k-q/2})\sigma^z\otimes\tau^z V(\mathbf{k+q/2})$. $W(\mathbf{k,q})=W^{\dag}(\mathbf{k,-q})$. Because of the inversion symmetry, we have $V(-\mathbf{k})=\sigma^x\otimes\tau^0 V(\mathbf{k})$, then $W(\mathbf{k,q})=-W(\mathbf{-k,-q})$. Using this relation, Eq.(\ref{chi}) can be simplified:
%\begin{widetext}
\begin{eqnarray}
\chi(\mathbf{q})&=&\frac{1}{N}\sum_{\mathbf{k}}\bigg[
\frac{W_{12\mathbf{k,q}}W_{21\mathbf{k,-q}}}{E_{h\mathbf{k-q/2}}+E_{h\mathbf{k+q/2}}}\nonumber\\
&&+\frac{2 {W_{14\mathbf{k,q}}W_{41\mathbf{k,-q}}}}{E_{h\mathbf{k-q/2}}+E_{l\mathbf{k+q/2}}}+\frac{W_{34\mathbf{k,-q}}W_{43\mathbf{k,q}}}{E_{l\mathbf{k-q/2}}+E_{l\mathbf{k+q/2}}}\bigg] \nonumber
\end{eqnarray}
%\end{widetext}

Note that both anti-ferromagnetic order and ferromagnetic order preserve translational symmetry on honeycomb lattice.
Peak at $\Gamma$ point indicts anti-ferromagnetic order because we have set $\phi_{A}=-\phi_{B}=\phi$.

\section{Mean field theory of $SU(2)$ slave rotor theory}\label{rotor}

\subsection{$Z_{2}$ ans\"atz}

The ans\"atz $ \eta^{\dag}_{AB}=a_{F}\sigma^{0}, \eta'^{\dag}_{AB,a}=a'_{F}\sigma^{a}, \eta_{AB}=a_{Z}\sigma^{0}, \eta'_{AB,a}=a'_{Z}\sigma^{a}$
 breaks the $SU(2)$ gauge symmetry to $Z_{2}$.
As this ans\"atz preserves time reversal symmetry, it works for large $U$. In this case, the effective Lagrangian reads:
%\begin{widetext}
\begin{eqnarray}
\mathcal{L}=&&\sum_{i\in A,B}(f^{\dag}_{i\sigma}\partial_{\tau}f_{i\sigma})+
\frac{3}{2U\alpha}\sum_{i\in A,B}(\partial_{\tau}z^{\ast}_{\alpha}\partial_{\tau}z_{\alpha})\nonumber\\
&&-\lambda\sum_{i\in A,B}(z^{\ast}_{\alpha}z_{\alpha}-1)+6tNa^{0}_{F}a^{0}_{Z}+6t'Na^{'}_{F}a^{'}_{Z}\nonumber\\
&&-t\sum_{\langle A,B\rangle}a_{F}f^{\dag}_{A\sigma}f_{B\sigma}
-t'\sum_{\langle A,B\rangle_{x}}a^{'}_{F}\sigma f^{\dag}_{A\sigma}f^{\dag}_{B\sigma}\nonumber\\
&&-t'\sum_{\langle A,B\rangle_{y}}a^{'}_{F} f^{\dag}_{A\sigma}f^{\dag}_{B\sigma}
-t'\sum_{\langle A,B\rangle_{z}}a^{'}_{F}\sigma f^{\dag}_{A\sigma}f_{B\sigma}\nonumber\\
&&-t\sum_{\langle A,B\rangle}a_{Z}z^{\dag}_{A\alpha}z_{B\alpha}
+t'\sum_{\langle A,B\rangle_{x}}a^{'}_{Z} z^{\dag}_{A\alpha}z^{\dag}_{B\beta}\nonumber\\
&&-t'\sum_{\langle A,B\rangle_{y}}i a^{'}_{Z} z^{\dag}_{A\alpha}z^{\dag}_{B\beta}
-t'\sum_{\langle A,B\rangle_{z}}a^{'}_{Z}\alpha z^{\dag}_{A\alpha}z_{B\alpha}\nonumber%+h.c.
\end{eqnarray}
%\end{widetext}

Let $t_{\uparrow}=-ta_{F}f-t'a'_{F}$, $t_{\downarrow}=-ta_{F}f+t'a'_{F}$, $\Delta_{\uparrow}=-t' a'_{F}(e^{i k_1}+e^{i k_2})$ and $\Delta_{\downarrow}=-t' a'_{F}(e^{i k_1}-e^{i k_2})$, then eigenvalues of spinons are $\varepsilon_{F}=\pm\frac{1}{2}|t_{\sigma}\pm\Delta_{\sigma}|$, of rotors are  $(\varepsilon_{Z}-\lambda)^{2}=t^2a^{2}_{Z} ff^{\ast}+3t^{'2}a^{'2}_{Z}\pm\sqrt{(t^2a^{2}_{Z} ff^{\ast}+3t^{'2}a^{'2}_{Z})^2-|t^{'2}a^{'2}_{Z}(g_{+}g_{-}+1)-t^2a^{2}_{Z}f^2|^2}$. 
If $a'_{Z}t'<2.8ta_{Z}$, the minimal of rotor eigenvalues is at $\Gamma$ point. In the following we assume $a'_{Z}t'<2.8ta_{Z}$ and we  find this condition is satisfied.
The rotor condensed part is: $-3ta_{Z}N(z^{\ast}_{A1}z_{B1}+z^{\ast}_{A2}z_{B2})+t'a'_{Z}(1+i)(z_{A1}z_{B2}+z_{A2}z_{B1})-t'a'_{Z}(z^{\ast}_{A1}z_{B1}-z^{\ast}_{A2}z_{B2})+\lambda Nz^{\ast}_{A/B\alpha}z_{A/B\alpha}/2+h.c.$ and $\lambda_{min}=3ta_{Z}+\sqrt{3}t'a'_{Z}$. Let $\sum_{i\in A,B}z^{\ast}_{i\alpha}z_{i\alpha}=2z^2$, then $z_{A2}$=$z_{B2}$=$2z/\sqrt{6+2\sqrt{3}}$, $z_{B1}$=$z_{A1}$=$(i-1)(1+\sqrt{3})z/\sqrt{6+2\sqrt{3}}$, so the rotor condensed part becomes: $-6ta_{Z}Nz^2-2\sqrt{3}t'a'_{Z}Nz^2+2\lambda Nz^2$. The free energy is($U'=2\alpha U/3$):
%\begin{widetext}
\begin{eqnarray}\label{fe}
F&=&-T\sum_{\mathbf{k},i}\ln{(1+e^{-\beta \varepsilon_{F,i}})}\nonumber\\
&&+T\sum_{\omega_{n}}\sum_{\mathbf{k},i}\ln{(\frac{3\omega^2_{n}}{2U}+\varepsilon_{Z,i})}+\mathrm{const}\nonumber\\
&=&-T\sum_{\mathbf{k},i}\ln{(1+e^{-\beta \varepsilon_{F,i}})}
+\sum_{\mathbf{k},i}\sqrt{U'\varepsilon_{Z,i}}\nonumber\\
&&+2T\sum_{\mathbf{k},i}\ln{(1-e^{-\beta\sqrt{U'\varepsilon_{Z,i}}})}+\mathrm{const}
 \end{eqnarray}
% \end{widetext}
 where the constant term is  $-2N\lambda+6tNa_{F}a_{Z}+6t'Na'_{F}a'_{Z}-6ta_{Z}Nz^2-2\sqrt{3}t'a'_{Z}Nz^2+2\lambda Nz^2$ 
and the second term  in the last line is zero point energies of relativistic rotors. Taking derivatives with respect to the parameters we get the following self-consistent equations:

%\begin{widetext}
\begin{eqnarray}\label{mfez2}
\frac{\partial F}{\partial a_{F}}=&&\sum_{\mathbf{k},i}n_{f}(\varepsilon_{F,i})
\frac{\partial \varepsilon_{F,i}}{\partial a_{F}}+6tNa_{Z}=0 \nonumber\\
\frac{\partial F}{\partial a^{z}_{F}}=&&\sum_{\mathbf{k},i}n_{f}(\varepsilon_{F,i})
\frac{\partial \varepsilon_{F,i}}{\partial a^{'}_{F}}+6t'Na^{'}_{Z}=0 \nonumber\\
\frac{\partial F}{\partial a_{Z}}=&&\sum_{\mathbf{k},i}\frac{\sqrt{U'}}{2\sqrt{\varepsilon_{Z,i}}}
\coth{(\frac{\beta\sqrt{U'\varepsilon_{Z,i}}}{2})}
\frac{\partial \varepsilon_{Z,i}}{\partial a_{Z}}\nonumber\\
~~~~~&&+6tN(a_{F}-z^2)=0 \\
\frac{\partial F}{\partial a^{'}_{Z}}=&&\sum_{\mathbf{k},i}\frac{\sqrt{U'}}{2\sqrt{\varepsilon_{Z,i}}}
\coth{(\frac{\beta\sqrt{U'\varepsilon_{Z,i}}}{2})}
\frac{\partial \varepsilon_{Z,i}}{\partial a^{'}_{Z}}\nonumber\\
&&+6t'N(a^{'}_{F}-z^2/\sqrt{3})=0 \nonumber\\
\frac{\partial F}{\partial \lambda}=&&\sum_{\mathbf{k},i}\frac{\sqrt{U'}}{2\sqrt{\varepsilon_{Z,i}}}
\coth{(\frac{\beta\sqrt{U'\varepsilon_{Z,i}}}{2})}
-2N+2Nz^2=0\nonumber
\end{eqnarray}
%\end{widetext}

Solving them numerically, we also find two critical lines.  When the interaction is smaller than $U_{c1}$ we predict a $p-wave$ super-conducting phase which is not reliable. When $U>U_{c1}$ we get spin liquid phase, in both cases the spinons are gapless. 

\begin{figure}[htb]
%\centering
\includegraphics[width=0.45\textwidth]{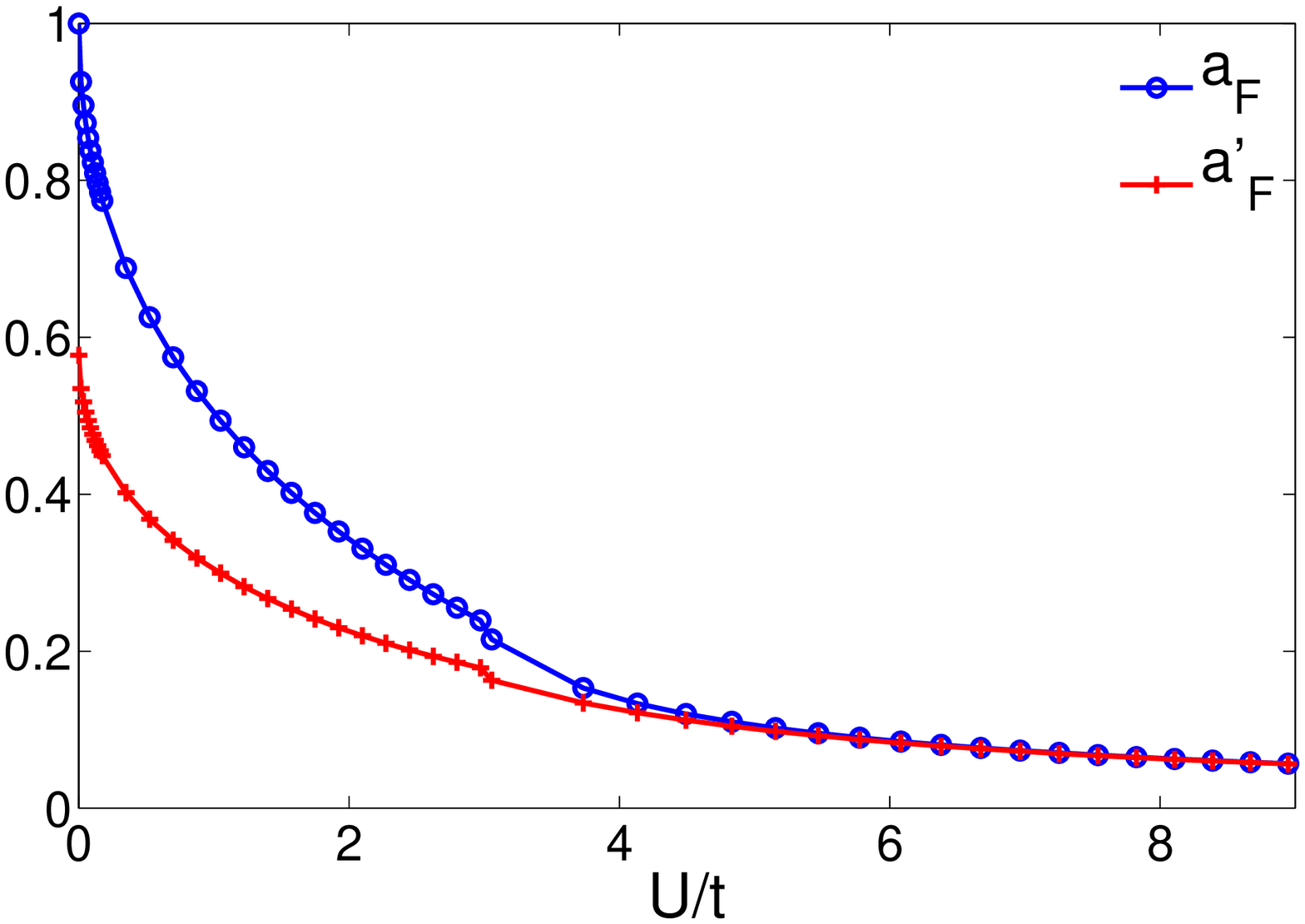}
\includegraphics[width=0.45\textwidth]{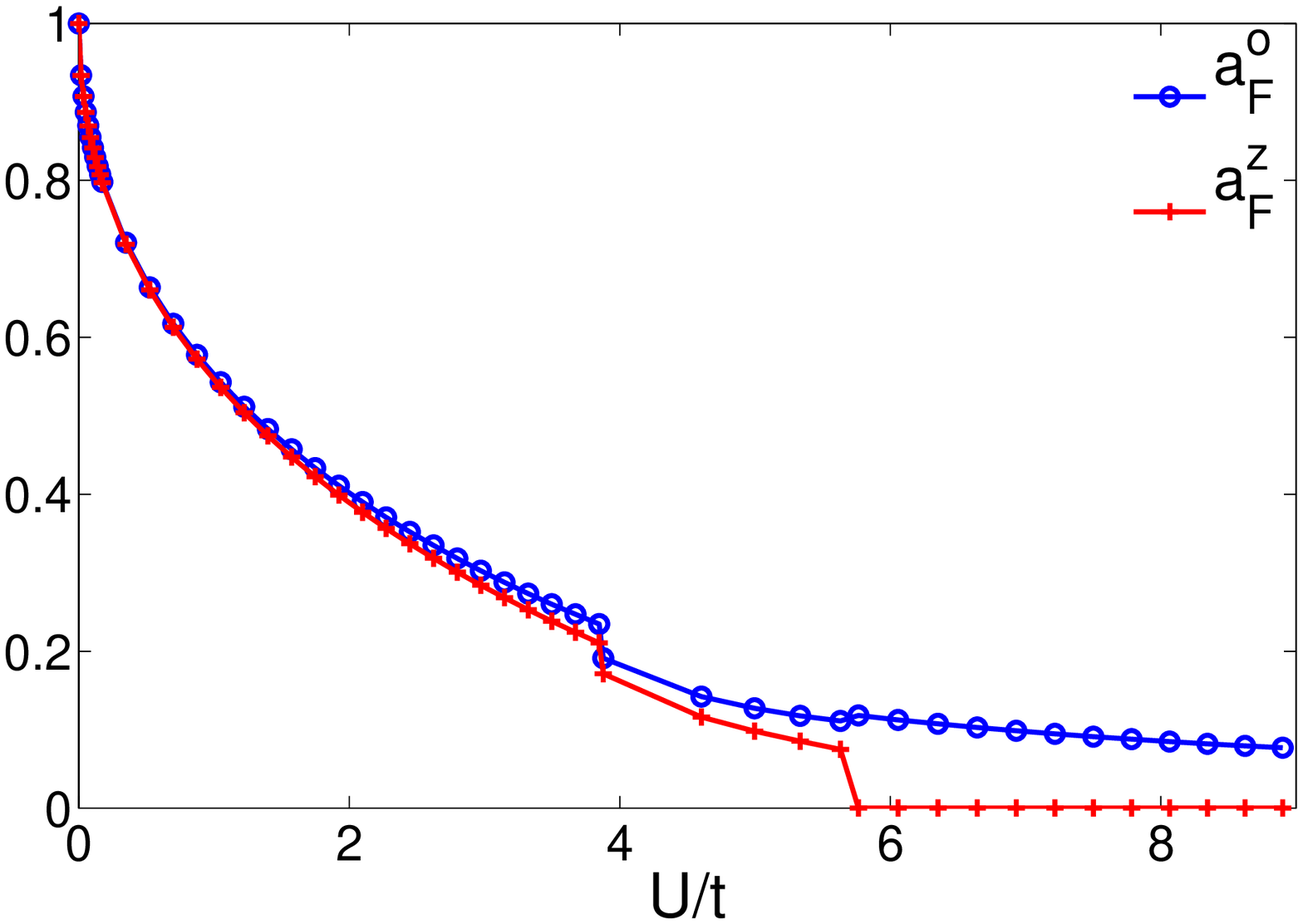}
%\centerline{(a)~~~~~~~~~~~~~~~~~~~~~~~~~~~~~~~~~~~~~~~~~~~~~~~~~~~~~~~~~~~~~~~~~~~~~~(b)}
\caption{(color online) Upper: $a_{F}$ and $a'_{F}$ for $Z_{2}$ ans\"atz. at $t=t'=1$. $a'_{F}/a_{F}\rightarrow1$ when $U$ increases.
Lower: $a^0_{F}$ and $a^z_{F}$ for $U(1)$ ans\"atz at $t=t'=1$. There are two phase transitions at $U_{c1}\approx 3.9t$ and
$U_{c2}\approx 5.8t$.}
\label{fig:af}
\end{figure}

In the spin liquid regime, when $t'/t\lesssim 0.91$ and $U>U_{c2}$, we get an $SU(2)$ spin liquid with $a'_{F}=a'_{Z}=0$. Otherwise it's a $Z_{2}$ spin liquid (See, e.g., Fig.\ref{fig:af}(a)).  

We study the large $U$ limit of $Z_{2}$ ans\"atz at $t'=t=1$ carefully. From Eq.(\ref{mfez2}), we know that in the large $U$ limit the effective chemical potential of rotors is $\lambda=U'$. Writing 
%\begin{widetext}
$\varepsilon_{F\sigma\pm}=\frac{1}{2}|t_{\sigma}\pm\Delta_{\sigma}|$, and  $\varepsilon_{\pm}=\big[t^2a^{2}_{Z} ff^{\ast}+3t^{'2}a^{'2}_{Z}\pm\big((t^2a^{2}_{Z} ff^{\ast}+3t^{'2}a^{'2}_{Z})^2-|t^{'2}a^{'2}_{Z}(g_{+}g_{-}+1)-t^2a^{2}_{Z}f^2|^2\big)^{1/2}\big]^{1/2}$,
% \end{widetext}
then in the large $U$ limit  Eq.(\ref{mfez2}) becomes:

\begin{eqnarray}\label{deta}
&&a_{Z}=\frac{1}{6tN}\sum_{\mathbf{k}}
\frac{\partial \varepsilon_{F,\sigma\pm}}{\partial a_{F}},~~ a'_{Z}=\frac{1}{6t'N}\sum_{\mathbf{k}}
\frac{\partial \varepsilon_{F,\sigma\pm}}{\partial a'_{F}} \\
&&
a_{F}=\frac{1}{24U'tN}\sum_{\mathbf{k}}
\frac{\partial \varepsilon^2_{Z,\pm}}{\partial a_{Z}},~~
a'_{F}=\frac{1}{24U't'N}\sum_{\mathbf{k}}
\frac{\partial \varepsilon^2_{Z,\pm}}{\partial a'_{Z}}\nonumber,
\end{eqnarray}

%\begin{minipage}{0.45\textwidth}
%\end{minipage}

If $t'=t$ and $U$ is large enough, $a'_{F}=a_{F}$(See, Fig.\ref{fig:af}(a)). Then the spinon dispersion becomes:  $\varepsilon_{F\uparrow+}=ta_{F}|f|$ and $\varepsilon_{F\uparrow-}=\varepsilon_{F\downarrow\pm}=ta_{F}$. We find two gapless dispersing bands and six gapped flat bands, which matches the exact solution of Kitaev model. 
In order to be consistent with this velocity, the mean field parameter $a_{F}$
 in the mean field spinon dispersion needs to be rescaled to $a_F=\frac{J}{16t}$\cite{kitaev2006,bn}. Solving Eq.\ref{deta} we find $a_{F}U'=0.1906t$, which gives $\alpha=0.572$. For $t'\ne t$, $a_{F}\ne a'_{F}$, the gapped flat bands acquire a weak dispersion\cite{ybk}.
 
\subsection{$U(1)$ ans\"atz}

The ans\"atz $\eta^{\dag}_{AB}=a^0_F\sigma^0, \eta'^{\dag}_{AB,a}=a^z_F\sigma^z,\eta_{AB}=a^0_Z\sigma^0, \eta'_{AB,a}=a^z_Z\sigma^z$
 breaks the $SU(2)$ gauge symmetry to $U(1)$. The spinon Hamiltonian has the same form as the noninteracting electron Hamiltonian with normalized hopping in this ans\"atz. We expect it is applicable to a small $U$. The symmetry of this ans\"atz is the same as the original model, e.g., time reversal symmetry is broken. As time reversal symmetry restores in the large $U$ limit, this ans\"atz does not work.
The effective Lagrangian is:
%\begin{widetext}
\begin{eqnarray}
\mathcal{L}&=&\sum_{i\in A,B}f^{\dag}_{i\sigma}\partial_{\tau}f_{i\sigma}+
\frac{3}{2U\alpha}\sum_{i\in A,B}(\partial_{\tau}z^{\ast}_{i\alpha}\partial_{\tau}z_{i\alpha})\nonumber\\
&&+\lambda\sum_{i\in A,B}(z^{\ast}_{i\alpha}z_{i\alpha}-1)+6tNa^{0}_{F}a^{0}_{Z}+6t'Na^z_{F}a^{z}_{Z}\nonumber\\
&&-ta^{0}_{F}\sum_{\langle A,B\rangle}f^{\dag}_{A\sigma}f_{B\sigma}
-t'a^{z}_{F}\sum_{\langle A,B\rangle_{a}}f^{\dag}_{A\sigma}\sigma^a_{\sigma\sigma'} f_{B\sigma'}\nonumber\\
&&-ta^{0}_{Z}\sum_{\langle A,B\rangle}z^{\dag}_{A\alpha}z_{B\alpha}
-t'a^{z}_{Z}\sum_{\langle A,B\rangle}z^{\dag}_{A\alpha}\sigma^{z}_{\alpha\beta}z_{B\beta}\nonumber
\end{eqnarray}
%\end{widetext}

Dispersion of spinons are 
$\varepsilon^{2}_{F}=t^2a^{02}_{F} ff^{\ast}+3t^{'2}a^{z2}_{F}\pm\big[(t^2a^{02}_{F} ff^{\ast}+3t^{'2}a^{z2}_{F})^2-|t^{'2}a^{z2}_{F}(g_{+}g_{-}+1)-t^2a^{02}_{F}f^2|^2\big]^{1/2}$. 
At Dirac points, the linear dispersion is proportional to $t'$ and  is not degenerate if $SU(2)$ symmetry is not restored. When, $a^z_F=0$, the $SU(2)$ gauge symmetry is restored, the dispersion becomes two-fold degenerate
%\begin{eqnarray}
$\varepsilon=\pm t |a^{0}_{F}| |f|$,
%\end{eqnarray} 
which is the same as the free Dirac fermion on the honeycomb lattice with the renormalized hopping $t|a^0_F|$.
The dispersion of rotors are $\varepsilon_{Z}=\lambda\pm|(ta^0_{Z}\pm t'a^z_{Z})f|$.
Rotors may condense at $\Gamma$ point, and we can write the condensed part explicitly($z_{A}=z_{B}$):
$-6Nta^{0}_{Z}(z^2_{1}+z^2_{2})-6Nt'a^z_{Z}(z^2_{1}-z^2_{2})+2N\lambda (z^2_{1}+z^2_{2})$.
The free energy is the same form as Eq.\ref{fe}
 with the constant term replaced by $-2N\lambda+6tNa^{0}_{F}a^{0}_{Z}+6t'Na^{z}_{F}a^{z}_{Z}-6Nta^{0}_{Z}(z^2_{1}+z^2_{2})-6Nt'a^z_{Z}(z^2_{1}-z^2_{2})+2N\lambda (z^2_{1}+z^2_{2})$. 
 Since $a^{0}_Z,a^{z}_{Z}>0$, we have $z_{2}=0$ and $\lambda_{min}=3(ta^{0}_{Z}+t'a^{z}_{Z})$. Taking derivatives with respect to the parameters we get the following self-consistent equations:

%\begin{widetext}
\begin{eqnarray}
\frac{\partial F}{\partial a^{0}_{F}}=&&\sum_{\mathbf{k},i}n_{f}(\varepsilon_{F,i})
\frac{\partial \varepsilon_{F,i}}{\partial a^{0}_{F}}+6tNa^{0}_{Z}=0 \nonumber\\
\frac{\partial F}{\partial a^{z}_{F}}=&&\sum_{\mathbf{k},i}n_{f}(\varepsilon_{F,i})
\frac{\partial \varepsilon_{F,i}}{\partial a^{z}_{F}}+6t'Na^{z}_{Z}=0 \nonumber\\
\frac{\partial F}{\partial a^{0}_{Z}}=&&\sum_{\mathbf{k},i}\frac{\sqrt{U'}}{2\sqrt{\varepsilon_{Z,i}}}
\coth{(\frac{\beta\sqrt{U'\varepsilon_{Z,i}}}{2})}
\frac{\partial \varepsilon_{Z,i}}{\partial a^{0}_{Z}}\nonumber\\
&&+6tN(a^{0}_{F}-z^2_{1})=0 \\
\frac{\partial F}{\partial a^{z}_{Z}}=&&\sum_{\mathbf{k},i}\frac{\sqrt{U'}}{2\sqrt{\varepsilon_{Z,i}}}
\coth{(\frac{\beta\sqrt{U'\varepsilon_{Z,i}}}{2})}
\frac{\partial \varepsilon_{Z,i}}{\partial a^{z}_{Z}}\nonumber\\
&&+6t'N(a^{z}_{F}-z^2_{1})=0 \nonumber\\
\frac{\partial F}{\partial \lambda}=&&\sum_{\mathbf{k},i}\frac{\sqrt{U'}}{2\sqrt{\varepsilon_{Z,i}}}
\coth{(\frac{\beta\sqrt{U'\varepsilon_{Z,i}}}{2})}
-2N+2Nz^2_{1}=0\nonumber
\end{eqnarray} 
%\end{widetext}

Solving these self-consistent equations numerically we find two critical lines: when the interaction is smaller than $\tilde U_{c1}$ we get a semi-metal phase, otherwise the rotors are gapped and we get a spin liquid phase. We find that $a^z_{F}$ and $a^z_{Z}$ are strongly suppressed when increasing $U$ and they vanish if $U>\tilde U_{c2}$. In this case we actually get an $SU(2)$ spin liquid, see Fig.\ref{fig:af}(b) . This can be understood in the following way: in the large $U$ limit there is an emergent time reversal symmetry, and there are two ways to recover this symmetry, that's, $a^z_{F}=a^z_{Z}=0$ or $a^0_{F}=a^0_{Z}=0$, because we are considering the $t'<t$ case, we get $a^z_{F}=a^z_{Z}=0$. This indicates that the $U(1)$ ans\"atz is not reliable in the large $U$ limit.

\subsection{$\alpha$ dependence of spin liquid phase diagram}\label{asl}

\begin{figure}
\includegraphics[width=0.45\textwidth]{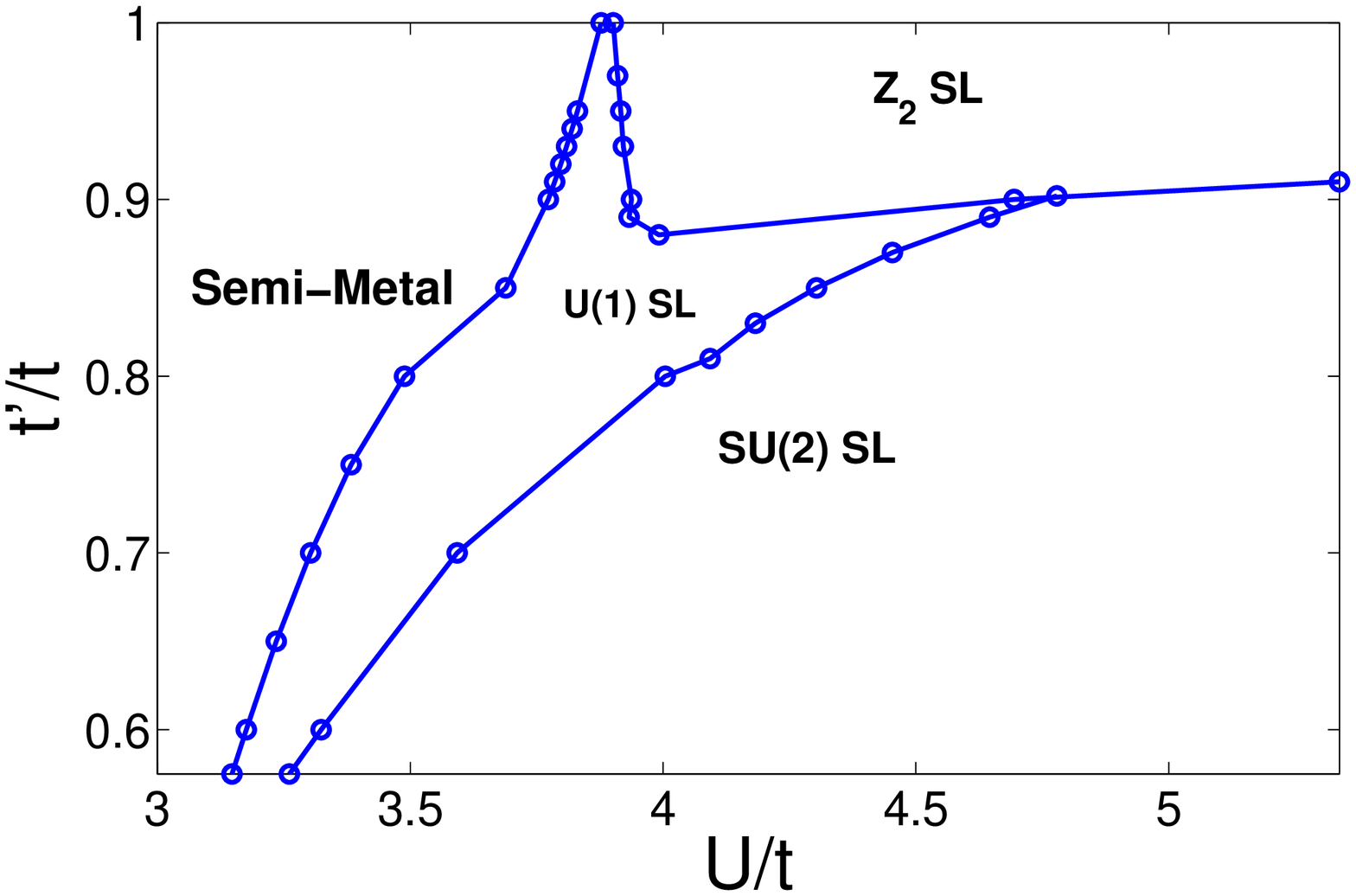}
\includegraphics[width=0.45\textwidth]{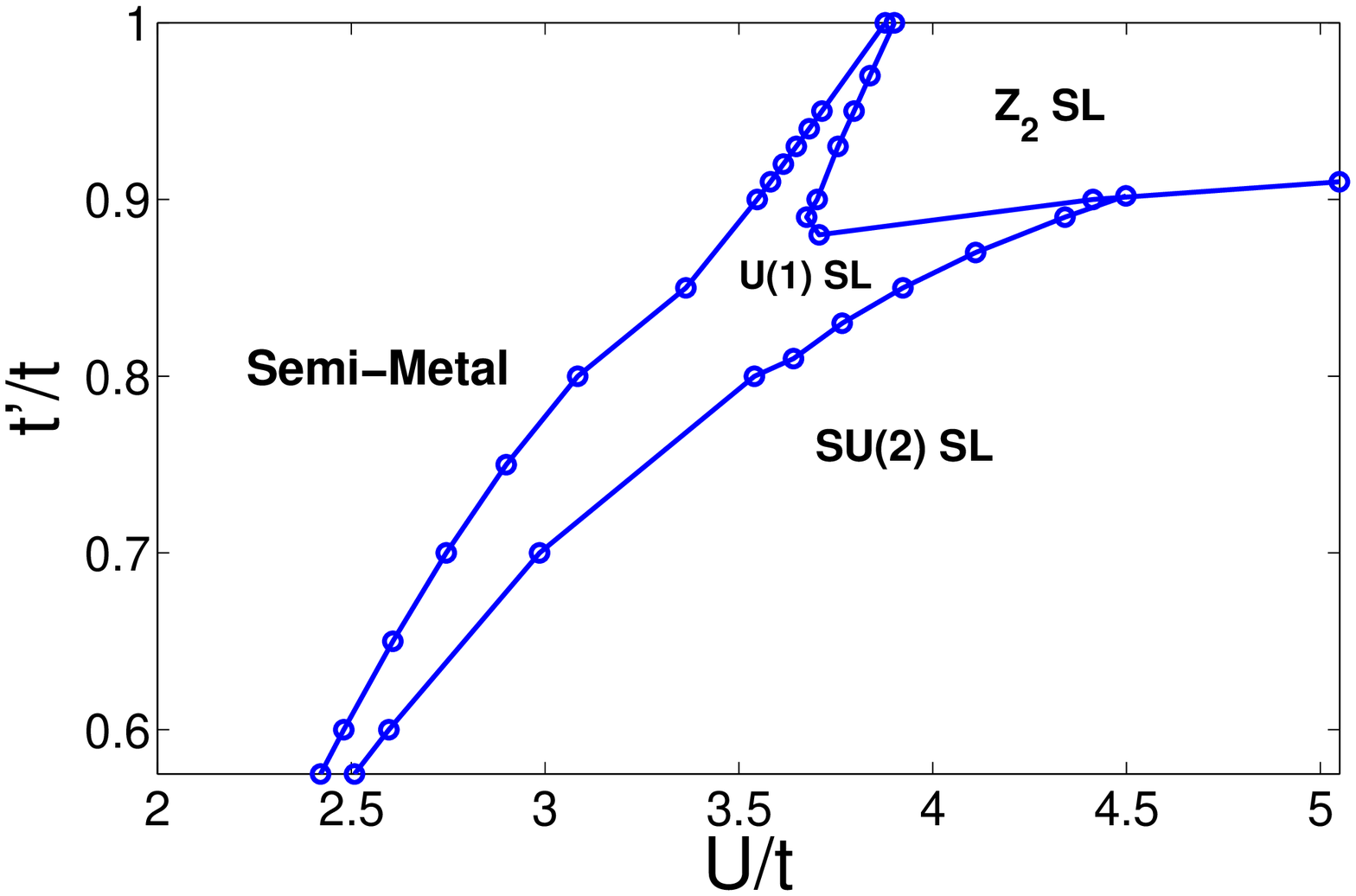}
\caption{(color online) $\alpha(t'/t=1)=0.572$ and varies linearly with $t'/t$. Upper: $\alpha(t'/t=0.575)=0.5$.
Lower: $\alpha(t'/t=1)=0.572$, $\alpha(t'/t=0.575)=0.65$.
}
\label{fig:difa}
\end{figure}

In the main text we choose $\alpha=0.572$ and get the spin liquid phase diagram Fig.\ref{fig:PhDMF}. We have explained why we choose the rescale parameter $\alpha$ as $t'/t$-independent. However, one may wonder\cite{question} what the phase diagram looks like if $\alpha$ is dependent on $t'/t$. The answer is $t'/t$-dependence of $\alpha$ doesn't change the topology of the spin liquid phase digram because $\alpha$ is only a rescale of the interaction. To show this we plot the phase diagram for different dependence of $\alpha$ on $t'/t$ (Fig.\ref{fig:difa}).

\end{document}